\documentclass{article}

\usepackage{arxiv}

\usepackage[utf8]{inputenc} 
\usepackage[T1]{fontenc}    
\usepackage{hyperref}       
\usepackage{url}            
\usepackage{booktabs}       
\usepackage{amsfonts}       
\usepackage{nicefrac}       
\usepackage{microtype}      
\usepackage{lipsum}
\usepackage{graphicx}
\usepackage{subcaption}
\usepackage{amsmath}
\usepackage{algorithm}
\usepackage{algorithmic}
\usepackage{booktabs}
\usepackage{color}
\usepackage{amssymb}
\usepackage{placeins}
\graphicspath{ {./images/} }

\title{BRP-GABLE: An Interpretable Rule-Based Prediction Framework for Multiple Outcome Types}

\author{%
Yulong Li\textsuperscript{1}
\quad
Ke Wan\textsuperscript{2}
\quad
Toshio Shimokawa\textsuperscript{3}\thanks{Corresponding author}
\quad
Kensuke Tanioka\textsuperscript{4}
\\[1.0em]%
\textsuperscript{1}Department of Biostatistics,\\
Graduate School of Medicine and Pharmaceutical Sciences,\\
Wakayama Medical University,\\
Wakayama, Japan
\\[0.6em]%
\textsuperscript{2}Department of Medical Data Science,\\
Wakayama Medical University,\\
Wakayama, Japan
\\[0.6em]%
\textsuperscript{3}Department of Biostatistics,\\
Wakayama Medical University,\\
Wakayama, Japan
\\[0.6em]%
\textsuperscript{4}Department of Biomedical Sciences and Informatics,\\
Doshisha University,\\
Kyoto, Japan
\\[0.8em]%
\textsuperscript{1}\texttt{yulongli123321@gmail.com}
\quad
\textsuperscript{2}\texttt{wane19911017@gmail.com}
\\
\textsuperscript{3}\texttt{toshibow2000@gmail.com}
\quad
\textsuperscript{4}\texttt{ktanioka@mail.doshisha.ac.jp}
\\[0.4em]%
}

\begin{document}
\maketitle
\begin{abstract}
Interpretable prediction models are important in biomedical research, where predictive accuracy must often be balanced against the ability to examine predictor-outcome relationships. Automatic Binary Logistic Estimation (ABLE) provides interpretable rule-based representations by constructing additive models using threshold rules. However, the original method is restricted to binary outcomes, and its greedy forward construction follows a search path, making candidate rules dependent on earlier selections and potentially excluding alternative structures. We extended ABLE to binary, continuous, and time-to-event outcomes using outcome-specific fitting criteria, yielding Generalized ABLE (GABLE). We then propose bootstrap rule pooling for GABLE (BRP-GABLE), which applies GABLE rule generation to bootstrap samples to explore alternative search paths, pool distinct rules, and perform a single global LASSO selection over the pooled rules and truncated linear terms. Unlike conventional bagging, bootstrap-specific models are not averaged, and resampling is used to diversify the candidate rule space before constructing a final model. Simulations showed that BRP-GABLE consistently improved predictive performance compared with GABLE and provided competitive prediction with relatively simple rule structures. Application to overall survival demonstrated its use for clinically interpretable prediction. BRP-GABLE provides a flexible framework for constructing sparse and interpretable predictive models across multiple outcome types.
\end{abstract}
\section{Introduction}

In clinical and biomedical data analysis, predictive models are often expected not only to achieve high predictive performance but also to provide interpretable representations of the relationships between predictors and outcomes. Clinical prediction models have been used to support diagnosis, prognosis, risk stratification, treatment decisions, and clinical study designs \cite{collins2015,steyerberg2013progress}. Machine learning methods have attracted considerable interest for the development of such models \cite{chen2017}. These include random forest (RF) \cite{breiman2001} and gradient boosting \cite{friedman2001}, which can flexibly capture nonlinear effects and interactions and achieve strong predictive performance. However, the prediction mechanisms of these ensemble models are often difficult to interpret directly, and post hoc explanations may not faithfully represent how fitted models generate predictions \cite{rudin2019}, potentially limiting clinical scrutiny and scientific interpretation \cite{kelly2019}. These considerations have motivated the development of prediction methods that combine modeling flexibility with intrinsically interpretable structures.

Among modeling approaches with directly interpretable structures, this study focuses on rule-based models that represent predictor--outcome relationships through explicit threshold conditions and their combinations. A well-known example is classification and regression trees (CART), which recursively partition the predictor space and can identify nonlinear effects and interactions without requiring them to be pre-specified \cite{breiman1984}. As an alternative rule-based construction, Tibshirani and LeBlanc proposed Automatic Binary Logistic Estimation (ABLE) for binary outcomes \cite{tibshirani1992}. ABLE constructs an additive model from binary basis functions defined by threshold conditions\cite{tibshirani1992}. Unlike CART, ABLE retains a basis function after it is split, allowing multiple main- and lower-order effects to coexist in the fitted model. The original study demonstrated that this construction could be more effective than CART in uncovering main effects and providing a simpler description of the data \cite{tibshirani1992}.

Despite these advantages, the original ABLE framework has two major limitations. First, it was developed primarily for binary outcomes, whereas continuous and time-to-event outcomes are common in biomedical research. Second, candidate rules are generated through a single greedy forward procedure. Therefore, a single run of ABLE explores only one possible search path and omits useful alternative rules. A related strategy for broadening the candidate search is used in RuleFit, which extracts a large collection of rules from multiple decision trees, combines them with truncated linear terms, and applies a  least absolute shrinkage and selection operator (LASSO)-based selection to obtain a sparse final model \cite{friedman2008,tibshirani1996}. However, each RuleFit rule is derived from a path within an individual tree. Therefore, a threshold main effect may need to be represented by multiple path-specific rules that inherit the conditions introduced by earlier splits. In contrast, the basis-retention property of ABLE enables multiple first-order threshold rules to be generated directly within an additive model. This structural difference motivates a strategy that preserves the rule construction properties of ABLE while exploring multiple search paths and performing global sparse selection over the resulting candidate rules.

To address these limitations, we first extended ABLE to binary, continuous, and time-to-event outcomes using outcome-specific fitting criteria and referred to the resulting procedure as Generalized ABLE (GABLE). We then propose Bootstrap Rule Pooling for Generalized ABLE (BRP-GABLE). BRP-GABLE applies the rule generation stage of GABLE independently to multiple bootstrap samples, pools the resulting distinct rules, combines them with truncated linear terms, and performs a single global LASSO selection. Bootstrap resampling is used to diversify the candidate rule set for a single greedy search procedure. The proposed construction preserves GABLE's ability to generate first-order and lower-order rule structures directly, while allowing bootstrap-specific forward searches to explore alternative candidate rules. Pooling these rules before a single global selection step reduces the dependence on any greedy search path and yields a single sparse final model.

The contributions of this study are twofold. First, we formulated GABLE by extending the rule-construction mechanism of ABLE to binary, continuous, and time-to-event outcomes. Second, we developed BRP-GABLE, which applies the rule-generation stage of GABLE to multiple bootstrap samples and performs global LASSO selection over the pooled rules and truncated linear terms.

The remainder of this paper is organized as follows: Section~2 describes the generalized ABLE and BRP-GABLE procedures. Section~3 presents the simulation study, Section~4 reports the clinical data application, and Section~5 discusses the principal findings and presents conclusions.

\section{Bootstrap Rule Pooling for Generalized ABLE}

We propose Bootstrap Rule Pooling for Generalized ABLE (BRP-GABLE), a bootstrap-based extension that expands the candidate rule set generated by a generalized ABLE. Bootstrap resampling was used to diversify the forward rule construction paths and explore alternative rule structures. This section presents the final BRP-GABLE model, followed by the generalized ABLE, bootstrap-specific rule generation, candidate-rule pooling, global LASSO selection, and interpretability measures.

\subsection{Overview of the proposed method}

We considered three outcome types: binary, continuous, and time-to-event. For the \(i\)-th observation, let
\(
\boldsymbol{x}_i=(x_{i1},\ldots,x_{ip})^\top\in\mathbb{R}^p
\)
denote a \(p\)-dimensional predictor vector. For binary and continuous outcomes, the training dataset is defined as
\(
\mathcal{D}_n
=
\left\{
(\boldsymbol{x}_i,y_i)
\right\}_{i=1}^{n},
\)
where \(y_i\in\{0,1\}\) for a binary outcome and
\(
y_i\in\mathbb{R}
\) 
for a continuous outcome. For a time-to-event outcome, the training dataset is defined as follows:
\(
\mathcal{D}_n
=
\left\{
(\boldsymbol{x}_i,t_i,\delta_i)
\right\}_{i=1}^{n},
\)
where \(t_i\) is the observed time and
\(
\delta_i\in\{0,1\}
\) is the event indicator, with
\(
\delta_i=1
\) indicating an observed event and
\(
\delta_i=0
\) 
indicating censoring.

\begin{figure}[ht]
    \centering
    \includegraphics[width=1\linewidth]{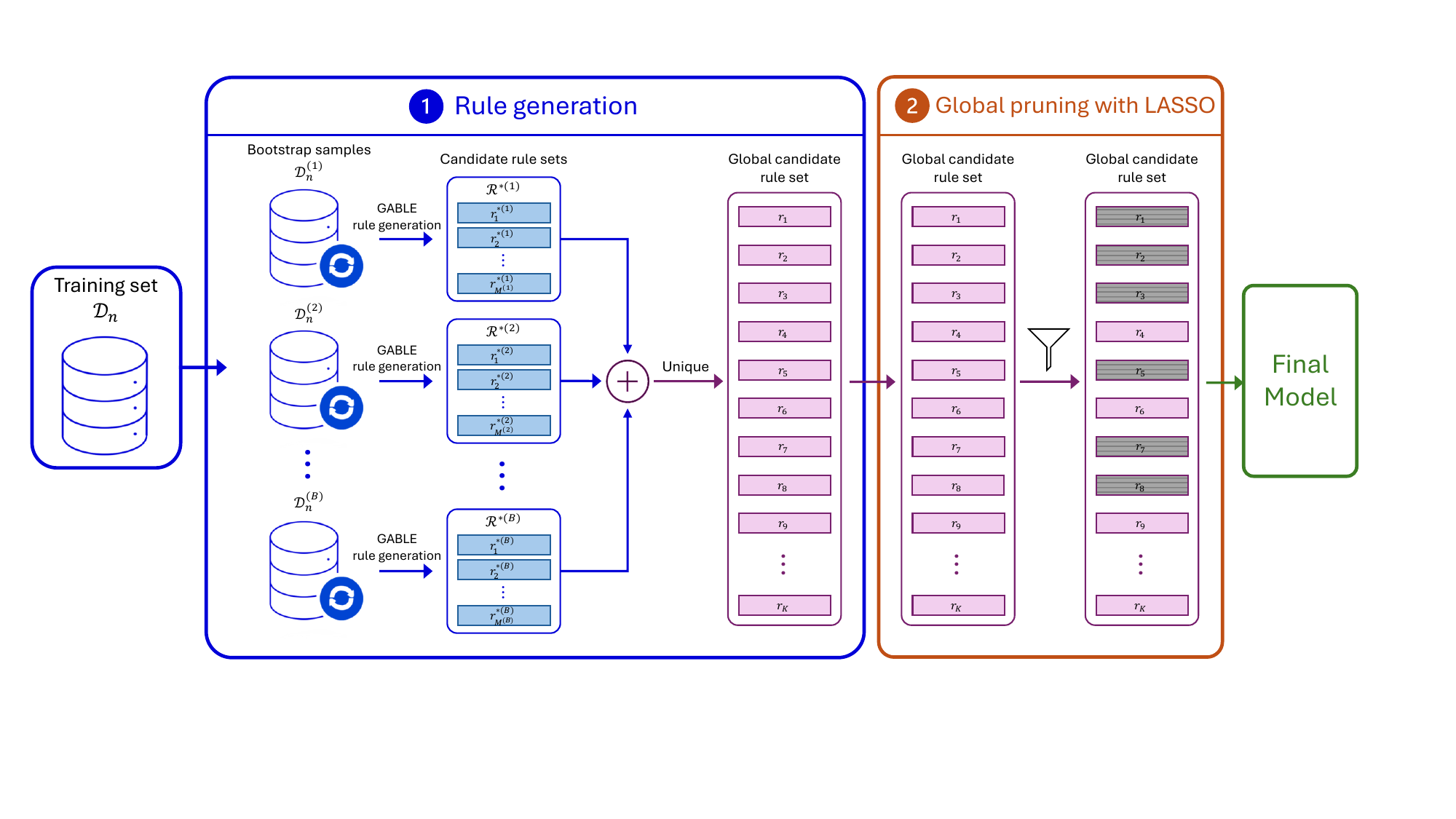}
    \caption{Overall workflow of BRP-GABLE.}
    \label{fig:workflow}
\end{figure}

BRP-GABLE consists of two stages. First, for each bootstrap sample, a generalized ABLE was used to construct rules according to an outcome-specific fitting criterion, and the resulting rules were pooled after exact duplicates were removed. Second, pooled rules and truncated linear terms were jointly selected using an outcome-specific global LASSO procedure. The overall workflow is illustrated in Figure~\ref{fig:workflow}.

Generalized ABLE was applied independently to each of the \(B\) bootstrap samples and the binary rule basis functions generated across these runs were pooled before exact duplicate rules were removed. Let
\[
\mathcal{R}
=
\{r_1,\ldots,r_K\}
\]
denote the resulting global candidate rule set, where each \(r_k(\boldsymbol{x})\in\{0,1\}\) represents a distinct rule-based function generated in at least one bootstrap sample and \(K\) is the total number of unique rules. A rule generated from multiple bootstrap samples appears only once in \(\mathcal{R}\) and receives a single coefficient in the final model. Detailed rule pooling and duplicate removal procedures are described in Section~2.3.2.

The final linear predictor is defined as follows:
\begin{equation}
\eta(\boldsymbol{x})
=
\begin{cases}
\displaystyle
\beta_0
+
\sum_{k=1}^{K}\beta_k r_k(\boldsymbol{x})
+
\sum_{j=1}^{p}\alpha_j l_j(x_j),
&
\text{(binary or continuous)},
\\[4mm]
\displaystyle
\sum_{k=1}^{K}\beta_k r_k(\boldsymbol{x})
+
\sum_{j=1}^{p}\alpha_j l_j(x_j),
&
\text{(time-to-event)}.
\end{cases}
\label{eq:final_model}
\end{equation}
Here, \(r_k(\boldsymbol{x})\) denotes the \(k\)-th rule from the global set \(\mathcal R\) of unique candidate rules and \(l_j(x_j)\) denotes the linear term of the \(j\)-th predictor, which is described in detail in Equation~\eqref{eq:truncated linear term}. Coefficients \(\beta_k\) and \(\alpha_j\) are the final coefficients associated with the retained rule and linear terms, which are obtained through outcome-specific LASSO pruning. Specifically, the intercept is included for binary and continuous outcomes but is omitted for time-to-event outcomes under the Cox proportional hazards formulation.

In addition to the rule-based predictors, the original predictors were incorporated as linear terms to capture global linear trends that may not be efficiently represented by rules alone. Following Friedman and Popescu~\cite{friedman2008}, each linear term is defined by a truncated version of the original predictor
\begin{equation}
l_j(x_j)
=
\min\left\{
u_j^{+},
\max\left(u_j^{-},x_j\right)
\right\},
\label{eq:truncated linear term}
\end{equation}
where \(u_j^{-}\) and \(u_j^{+}\) denote the lower and upper truncation points of the \(j\)-th predictor, respectively. In this study, these truncation points were determined from the training data using the 2.5th and 97.5th percentiles of each predictor.

\subsection{Generalized ABLE model}
The original ABLE method was developed for binary outcomes and consists of a forward-rule construction stage followed by backward hierarchical pruning \cite{tibshirani1992}. We generalized its rule construction mechanism to binary, continuous, and time-to-event outcomes by retaining the same forward construction procedure, while changing the fitting criterion according to the outcome type. We refer to the resulting outcome-generalized procedure as Generalized ABLE (GABLE).

Within BRP-GABLE, GABLE is used as a bootstrap-specific rule-generation procedure. The coefficients estimated during this stage were used solely to evaluate candidate rule extensions and were not included in the final model. Instead, only the generated rule functions were retained, pooled across bootstrap samples, and subsequently subjected to a single global LASSO selection.

To describe the model used during this rule-generation stage, the GABLE predictor is expressed as follows:
\begin{equation}
\eta^\ast(\boldsymbol{x})
=
\begin{cases}
\displaystyle
\beta_0^\ast
+
\sum_{m=1}^{M}\beta_m^\ast r_m^\ast(\boldsymbol{x})
,
&
\text{(binary or continuous)},
\\[4mm]
\displaystyle
\sum_{m=1}^{M}\beta_m^\ast r_m^\ast(\boldsymbol{x})
,
&
\text{(time-to-event)}.
\end{cases}
\label{eq:GABLE_model}
\end{equation}

Here, \(M\) is the number of generated rules, \(\beta^\ast_m\) is the coefficient of the \(m\)-th rule, and \(r_m^\ast(\boldsymbol{x})\in\{0,1\}\) is the \(m\)-th rule.

Each rule is defined as the combination of predictor-specific threshold
conditions, represented as the product of the indicator functions
\begin{equation}
r_m^\ast(\boldsymbol{x})
=
\prod_{d=1}^{d_m}
I\bigl(
s_{md}
\left(
x_{k_{md}}-c_{md}
\right)
>0
\bigr).
\end{equation}
Here, \(d_m\) denotes the degree of the \(m\)-th rule, which is defined as 
The number of threshold conditions in the rule 
\(k_{md}\)
identifies the predictor used in the \(d\)-th condition. 
\(c_{md}\) is the corresponding cutoff value, and \(s_{md}\in\{-1,1\}\)
is a sign parameter indicating the direction of the condition. Accordingly, \(s_{md}=1\) represents \(x_{k_{md}}>c_{md}\) and \(s_{md}=-1\) represents \(x_{k_{md}}\leq c_{md}\). For the boundary case \(x_{k_{md}}=c_{md}\), we adopted the convention that the observation is assigned to the left branch. The rule takes the value of one only when all its threshold conditions are satisfied.

\noindent\textbf{Outcome-specific fitting criterion.}

The forward rule construction mechanism is common to all three outcome types, whereas the candidate extensions are evaluated using an outcome-specific loss function. We used negative logistic log-likelihood for binary outcomes, squared-error loss for continuous outcomes, and negative Cox partial log-likelihood for time-to-event outcomes \cite{cox1972}. Let 
\(\boldsymbol{\eta}^\ast=(\eta_1^\ast,\ldots,\eta_n^\ast)^\top\),
where
\(\eta_i^\ast=\eta^\ast(\boldsymbol{x}_i)\).
We define
\begin{equation}
\mathcal{L}(\boldsymbol{\eta}^\ast)
=
\begin{cases}
\displaystyle
-\sum_{i=1}^{n}
\left[
y_i\log p_i^\ast
+
(1-y_i)\log(1-p_i^\ast)
\right],
&
\text{(binary)},
\\[4mm]
\displaystyle
\frac{1}{2}
\sum_{i=1}^{n}
(y_i-\eta_i^\ast)^2,
&
\text{(continuous)},
\\[4mm]
\displaystyle
-\sum_{i:\delta_i=1}
\left[
\eta_i^\ast
-
\log
\left\{
\sum_{i'\in\mathcal{Q}_i}
\exp(\eta_{i'}^\ast)
\right\}
\right],
&
\text{(time-to-event)}.
\end{cases}
\label{eq:outcome_loss}
\end{equation}
where
\[
p_i^\ast
=
\frac{1}{1+\exp(-\eta_i^\ast)}
\]
for binary outcomes and \(\mathcal{Q}_i\) is the risk set at time \(t_i\) for time-to-event outcomes.

\subsection{Rule generation and global pruning}

\subsubsection{Bootstrap rule generation}
BRP-GABLE extends the generalized ABLE procedure described above by repeatedly applying its rule-generation stage to bootstrap samples drawn from the original training set. For each bootstrap replication \(b=1,\ldots,B\), where \(B\) denotes the total number of bootstrap replications, the bootstrap sample is defined as
\(
\mathcal{D}_n^{(b)}
=
\left\{
\left(\boldsymbol{x}_i^{(b)},y_i^{(b)}\right)
\right\}_{i=1}^{n}
\)
for binary and continuous outcomes, and as
\(
\mathcal{D}_n^{(b)}
=
\left\{
\left(
\boldsymbol{x}_i^{(b)},
t_i^{(b)},
\delta_i^{(b)}
\right)
\right\}_{i=1}^{n}
\)
for time-to-event outcomes. In both cases, \(n\) observations are sampled with replacement from the corresponding training dataset \(\mathcal{D}_n\).

The purpose of the forward construction procedure is to generate a candidate rule model for a given training set that corresponds to one bootstrap sample in BRP-GABLE. Starting with a constant rule, the procedure sequentially expands the current model by adding a candidate rule extension that provides the largest improvement to the outcome-specific fitting criterion. The coefficients estimated during this procedure are used only to construct the rules; they are not carried over to the final BRP-GABLE model. Only the generated rule functions are retained as candidate predictors and are later pooled across bootstrap samples and pruned globally using LASSO.

\begin{algorithm}[htbp]
\caption{Forward construction procedure for the \(b\)-th bootstrap sample}
\label{alg:forward_generalized_able}
\begin{algorithmic}[1]
\STATE \textbf{Input:} Bootstrap sample \(\mathcal{D}_n^{(b)}\), maximum number of model terms \(M_{\max}\), maximum interaction degree \(d_{\max}\), improvement tolerance \(\varepsilon\)
\STATE Initialize \(r_0^{\ast(b)}(\boldsymbol{x}) \leftarrow 1,\quad M_{\mathrm{next}}^{(b)} \leftarrow 1,\quad Loss^\ast \leftarrow \infty,\quad d_0^{(b)} \leftarrow 0\)
\WHILE{\(M_{\mathrm{next}}^{(b)} \leq M_{\max}\)}
\STATE\(Loss_{\mathrm{old}}\leftarrow Loss^\ast,\quad(m^\ast,j^\ast,c^\ast)\leftarrow\varnothing\)
\FOR{\(m=0,\ldots,M_{\mathrm{next}}^{(b)}-1\)}
\IF{\(d^{(b)}_m < d_{\max}\)}
\FOR{\(j=1,\ldots,p\)}
\IF{\(x_j\) is not included in \(r_m^{\ast(b)}(\boldsymbol{x})\)}
\STATE Construct the admissible cutoff set
\(\mathcal{C}_{mj}^{(b)}\) from the observed values of \(x_j\)
\FOR{\(c \in \mathcal{C}_{mj}^{(b)}\)}
\STATE \(g \leftarrow
\sum_{a=0}^{M_{\mathrm{next}}^{(b)}-1}\beta_a^{\ast(b)}r_a^{\ast(b)}(\boldsymbol{x})+\beta^{\ast(b)}_{M_{\mathrm{next}}^{(b)}}r_m^{\ast(b)}(\boldsymbol{x})I(x_j \leq c)+\beta^{\ast(b)}_{M_{\mathrm{next}}^{(b)}+1}r_m^{\ast(b)}(\boldsymbol{x})I(x_j > c)\)
\STATE \(Loss \leftarrow
\min_{\beta^{\ast(b)}_0,\ldots,\beta^{\ast(b)}_{M_{\mathrm{next}}^{(b)}+1}}
\mathcal{L}^{(b)}(g)\)
\IF{\(Loss < Loss^\ast\)}
\STATE \(Loss^\ast \leftarrow Loss,\quad
(m^\ast,j^\ast,c^\ast)\leftarrow(m,j,c)\)
\ENDIF
\ENDFOR
\ENDIF
\ENDFOR
\ENDIF
\ENDFOR
\IF{\((m^\ast,j^\ast,c^\ast)=\varnothing
\ \OR\
Loss_{\mathrm{old}}-Loss^\ast\leq\varepsilon\)}
\STATE \textbf{break}
\ENDIF
\STATE \(r_{M_{\mathrm{next}}^{(b)}}^{\ast(b)}(\boldsymbol{x})
\leftarrow r_{m^\ast}^{\ast(b)}(\boldsymbol{x})I(x_{j^\ast}\leq c^\ast)\)
\STATE \(r_{M_{\mathrm{next}}^{(b)}+1}^{\ast(b)}(\boldsymbol{x})
\leftarrow r_{m^\ast}^{\ast(b)}(\boldsymbol{x})I(x_{j^\ast}>c^\ast)\)
\STATE
\(d_{M_{\mathrm{next}}^{(b)}}^{(b)}
\leftarrow d_{m^\ast}^{(b)}+1,\quad
d_{M_{\mathrm{next}}^{(b)}+1}^{(b)}
\leftarrow d_{m^\ast}^{(b)}+1\)
\STATE \(M_{\mathrm{next}}^{(b)} \leftarrow M_{\mathrm{next}}^{(b)}+2\)
\ENDWHILE
\STATE \textbf{Output:} Bootstrap-specific candidate rule set \(\mathcal{R}^{\ast(b)}=\{r_1^{\ast(b)},\ldots,r_{M_{\mathrm{next}}^{(b)}-1}^{\ast(b)}\}\)
\end{algorithmic}
\end{algorithm}

Algorithm~\ref{alg:forward_generalized_able} describes the forward construction procedure applied to the \(b\)-th bootstrap sample. For a temporary linear predictor \(g\) evaluated on the \(b\)-th bootstrap,
sample, define
\[
\mathcal{L}^{(b)}(g)
=
\mathcal{L}
\left(
\left\{
g(\boldsymbol{x}_i^{(b)})
\right\}_{i=1}^{n};
\mathcal{D}_n^{(b)}
\right),
\]
where \(\mathcal{L}\) denotes the outcome-specific loss defined in Eq.~\eqref{eq:outcome_loss}. In the following algorithm, \(Loss\) denotes the minimized loss of the candidate model currently being evaluated. At the beginning of each forward iteration, \(Loss^\ast\) denotes the minimized loss of the currently accepted model, which is updated whenever a candidate model with a smaller loss is identified.

For each predictor \(x_j\) not already included in \(r_m^{\ast(b)}(\boldsymbol{x})\), candidate cutoffs are constructed using only the observations for which the parent rule is active, that is, observations satisfying \(r_m^{\ast(b)}(\boldsymbol{x}_i^{(b)})=1\). A cutoff \(c\) is eligible only when both the resulting child rules contain at least \(n_{\min}\) observations.
\[
\sum_{i=1}^{n}
r_m^{\ast(b)}(\boldsymbol{x}_i^{(b)})
I(x_{ij}^{(b)}\leq c)
\geq n_{\min},
\qquad
\sum_{i=1}^{n}
r_m^{\ast(b)}(\boldsymbol{x}_i^{(b)})
I(x_{ij}^{(b)}>c)
\geq n_{\min},
\]
where
\(
n_{\min}
=
\left\lceil 0.05n\right\rceil.
\)
Candidate cutoffs are constructed from observations that satisfy the parent rule, and each resulting child rule is required to contain at least 5\% of the full bootstrap sample.

To reduce computational burden, not every eligible observed value was examined. The eligible values are ordered within the active observations of the parent rule and a centered, sparsely spaced set of cutoff candidates is selected using a rank interval of
\(
h=\left\lceil 0.20n\right\rceil.
\)
The grid is centered within the range permitted by the minimum support condition. Each target rank is mapped to the first eligible observed value whose cumulative count reaches that rank, and duplicated cutoff values are removed. The resulting parent- and predictor-specific candidate sets are denoted by \(\mathcal{C}_{mj}^{(b)}\). This rank-spaced screening strategy was motivated by the minimum-span concept used in MARS to avoid an excessively dense search for possible knot locations \cite{friedman1991}. However, the proposed procedure uses a fixed parent-conditional rank grid and is not identical to the original MARS minimum-span specification. Using the constant parent rule and a few tied values, setting \(h=\lceil0.20n\rceil\) typically produced approximately five candidate cutoffs per predictor. Fewer candidates may be available for higher-order parent rules because their active support is smaller.

For every combination of an expandable rule \(r_m^{\ast(b)}(\boldsymbol{x})\), an eligible predictor \(x_j\), and a cutoff \(c\in\mathcal{C}_{mj}^{(b)}\), the corresponding temporary model \(g\) is constructed as shown in Algorithm~\ref{alg:forward_generalized_able}. We re-estimate all coefficients in the temporary model by minimizing the bootstrap-specific outcome loss \(\mathcal{L}^{(b)}(g)\). The resulting minimized value \(Loss\) is compared with \(Loss^\ast\). Whenever a candidate produces a loss smaller than \(Loss^\ast\), the corresponding indices \((m^\ast,j^\ast,c^\ast)\) are updated.

After all admissible combinations of \(m\), \(j\), and \(c\) are examined, the reduction in loss is evaluated by comparing \(Loss^\ast\) with \(Loss_{\mathrm{old}}\), the loss stored at the beginning of the current forward iteration. If no admissible candidate is found, or the reduction in loss does not exceed the tolerance \(\varepsilon\), the forward construction terminates. Otherwise, the sibling pair corresponding to \((m^\ast,j^\ast,c^\ast)\) is added to the current rule set and the updated value of \(Loss^\ast\) is retained for the next forward iteration. Thus, the algorithm evaluates every cutoff in each constructed sparse candidate set \(\mathcal{C}_{mj}^{(b)}\) rather than every distinct observed value, and selects the extension that produces the greatest loss reduction.

The sibling-pair construction described above is used for predictors with more than two distinct values. For binary predictors coded as 0 and 1, a single rule of the form
\[
r_m^{\ast(b)}(\boldsymbol{x})I(x_j=1)
\]
was added following the original ABLE rule construction procedure. Algorithm~\ref{alg:forward_generalized_able} presents the sibling-pair construction for nonbinary predictors; the corresponding modification for a binary predictor is as described above.

The forward construction continues until the maximum number of model terms \(M_{\max}\) is reached, no admissible extension remains, or the best available extension fails to reduce the current loss by more than \(\varepsilon\), which was set to \(10^{-8}\) in this study. The output is a bootstrap-specific candidate rule set.
\[
\mathcal{R}^{\ast(b)}
=
\{r_1^{\ast(b)},\ldots,r_{M_{\mathrm{next}}^{(b)}-1}^{\ast(b)}\},
\]
excluding the constant rule, \(r_0^{\ast(b)}(\boldsymbol{x})\). We denote the
number of nonconstant rules generated from the \(b\)-th bootstrap sample by
\[
M^{(b)}
=
M_{\mathrm{next}}^{(b)}-1.
\]
The coefficients estimated during forward construction are used only to evaluate the candidate extensions and are not included in the final BRP-GABLE model. The final coefficients are estimated using the global LASSO procedure.

\noindent\textbf{Example.}
An illustrative example of the resulting rule structure is presented below. From the constant rule \(r_0^{\ast(b)}(\boldsymbol{x})=1\), the first split may generate a sibling pair such that
\[
r_1^{\ast(b)}(\boldsymbol{x})=I(x_1\leq c_1)
\quad\text{and}\quad
r_2^{\ast(b)}(\boldsymbol{x})=I(x_1>c_1).
\]
Subsequent splits can then be applied to one of the existing rules, yielding
higher-order rules, such as
\[
r_5^{\ast(b)}(\boldsymbol{x})
=
I(x_1>c_1)I(x_3\leq c_3)
\quad\text{and}\quad
r_6^{\ast(b)}(\boldsymbol{x})
=
I(x_1>c_1)I(x_3>c_3).
\]
Thus, more complex interaction rules are generated by sequentially appending indicator functions to previously generated rules. Applying this procedure to all the bootstrap samples yields bootstrap-specific candidate rule sets \(\mathcal{R}^{\ast(1)},\ldots,\mathcal{R}^{\ast(B)}\). These rule sets are then pooled, combined with the truncated linear terms, and pruned globally, as described in the following subsection.

\subsubsection{Rule pooling, duplicate removal, and outcome-specific global pruning}

After the bootstrap rule-generation procedure, BRP-GABLE obtains \(B\) bootstrap-specific candidate rule sets
\(
\mathcal{R}^{\ast(1)},\ldots,\mathcal{R}^{\ast(B)}.
\)
The rule-based functions contained in these sets are pooled and exact duplicate rules are removed. To identify exact duplicates, each rule is represented by its elementary conditions, including the predictor, cutoff value, and direction of inequality. The conditions are placed in a fixed order such that rules containing the same conditions are identified as identical, regardless of the order in which the conditions are generated. Two rules are treated as exact duplicates if all their elementary conditions are identical.

One representative from each set of duplicates is retained from the pooled candidate rules. The resulting global candidate rule set is denoted by
\[
\mathcal{R}
=
\left\{
r_m^{\ast(b)}
\;\middle|\;
b\in\{1,\ldots,B\},
\quad
m\in\{1,\ldots,M^{(b)}\}
\right\}
=
\{r_1,\ldots,r_K\}.
\label{eq:global_rule_set}
\]
where \(K=|\mathcal{R}|\) is the total number of distinct candidate rules. Every rule generated in bootstrap-specific candidate sets is either represented by one element of \(\mathcal{R}\) or is an exact duplicate of one of its elements. Thus, if the same rule is generated from multiple bootstrap samples, it is retained only once.

Rules containing different predictors, cutoff values, or inequality directions are retained as distinct candidates even if they produce identical activation values for all observations in the available data.

The resulting candidate rules are combined with the truncated linear terms introduced in Section~2.1 to form the complete linear predictor given in Equation~\eqref{eq:final_model}. Let
\(\boldsymbol{\beta}=(\beta_1,\ldots,\beta_K)^\top\) and
\(\boldsymbol{\alpha}=(\alpha_1,\ldots,\alpha_p)^\top\)
denote the coefficients of the candidate rule terms and truncated linear terms, respectively. Before [A8.1]LASSO fitting, each rule indicator and truncated linear term is scaled by the standard deviation estimated from the training data. Using the outcome-specific losses
\(\mathcal{L}(\boldsymbol{\eta})\) defined in Equation~\eqref{eq:outcome_loss} and the linear predictor defined in Equation~\eqref{eq:final_model}, the final coefficients are estimated by solving
\begin{equation}
\left(
\widehat{\beta}_0,
\widehat{\boldsymbol{\beta}},
\widehat{\boldsymbol{\alpha}}
\right)
=
\arg\min_{\beta_0,\boldsymbol{\beta},\boldsymbol{\alpha}}
\left[
\frac{1}{c_y}
\mathcal{L}(\boldsymbol{\eta})
+
\lambda
\left(
\sum_{k=1}^{K}|\beta_k|
+
\sum_{j=1}^{p}|\alpha_j|
\right)
\right],
\label{eq:global_lasso}
\end{equation}
where
\[
c_y
=
\begin{cases}
n,
&
\text{(binary or continuous)},
\\[1mm]
n_{\mathrm{event}},
&
\text{(time-to-event)},
\end{cases}
\qquad
n_{\mathrm{event}}
=
\sum_{i=1}^{n}\delta_i.
\]
For time-to-event outcomes, the intercept \(\beta_0\) is omitted from both the linear predictor and optimization problem under the Cox proportional hazards formulation. For binary and continuous outcomes, the intercept is included but not penalized.

The scaling factor \(c_y\) expresses each outcome-specific loss on an average per observation or event scale. Depending on the outcome type, Equation~\eqref{eq:global_lasso} corresponds to LASSO-penalized logistic regression, linear regression, or Cox proportional hazards regression. In all three settings, the regularization parameter \(\lambda\) is selected using 10-fold cross-validation and the one-standard-error rule. Only the coefficients estimated in this global selection step are used in the final BRP-GABLE model; the bootstrap-specific coefficients used during the forward rule construction are not carried forward.

\subsection{Model interpretation tools}
\noindent\textbf{Term importance.}

To facilitate the interpretation of the final BRP-GABLE model, we defined the global importance of the retained rule and linear terms following the interpretation strategy of RuleFit \cite{friedman2008}.

For a retained rule \(r_k(\boldsymbol{x})\), its global importance is defined as follows:
\[
I_k^{\mathrm{rule}}
=
|\widehat{\beta}_k|
\sqrt{{\pi}_k(1-{\pi}_k)},
\]
where \(\widehat{\beta}_k\) denotes the coefficient of the rule on the original unnormalized rule scale and
\[
{\pi}_k
=
\frac{1}{n}
\sum_{i=1}^{n}
r_k(\boldsymbol{x}_i)
\]
denotes the empirical support in the training data.

For a retained linear term \(l_j(x_j)\), its global importance is
defined as
\[
I_j^{\mathrm{lin}}
=
\left|\widehat{\alpha}_j\right|
\operatorname{sd}\bigl(l_j(x_j)\bigr),
\]
where \(\widehat{\alpha}_j\) denotes the estimated coefficient associated with \(l_j(x_j)\) in the final model and
\(\operatorname{sd}\bigl(l_j(x_j)\bigr)\) denotes the empirical standard deviation of
\(\{l_j(x_{ij})\}_{i=1}^{n}\) over the data used to fit the final model.

For presentation, the term importance values may be rescaled such that the largest value is set to 100.

\noindent\textbf{Accumulated local effects.}

To visualize how the fitted prediction changes over the observed range of a continuous predictor, we used first-order accumulated local effects (ALE) \cite{apley2020}. ALE divides the observed range of a predictor into intervals, estimates the average local change in the fitted prediction within each interval, and accumulates these local changes across the predictor range.

Let \(\widehat{\eta}(\boldsymbol{x})\) denote the fitted linear predictor. The observed range of the predictor \(x_j\) is divided into \(L\) intervals with boundaries:
\[
q_{j,0}<q_{j,1}<\cdots<q_{j,L}.
\]
Let \(\boldsymbol{x}_{i,-j}\) denote the observed values of all predictors other than \(x_j\) for the \(i\)-th observation. For any value \(z\) of predictor \(x_j\), \(\widehat{\eta}(z,\boldsymbol{x}_{i,-j})\) denotes the fitted linear predictor evaluated by setting \(x_j=z\) while holding all other predictors at their observed values \(\boldsymbol{x}_{i,-j}\).
Let \(\mathcal{N}_{j\ell}\) denote the set of observations falling in the \(\ell\)-th interval and let \(n_{j\ell}=|\mathcal{N}_{j\ell}|\). The average local effect within the interval was calculated as follows:
\[
\widehat{\Delta}_{j\ell}
=
\frac{1}{n_{j\ell}}
\sum_{i\in\mathcal{N}_{j\ell}}
\left[
\widehat{\eta}
\left(
q_{j,\ell},\boldsymbol{x}_{i,-j}
\right)
-
\widehat{\eta}
\left(
q_{j,\ell-1},\boldsymbol{x}_{i,-j}
\right)
\right].
\label{eq:ale_local_effect}
\]

Let \(\nu_j(x_j)\in\{1,\ldots,L\}\) denote the index of the interval containing \(x_j\). The ALE function is obtained by accumulating the average local effects up to that interval and then centering the resulting function as follows:
\[
\widehat{f}_{j,\mathrm{ALE}}(x_j)
=
\sum_{\ell=1}^{\nu_j(x_j)}
\widehat{\Delta}_{j\ell}
-
\frac{1}{n}
\sum_{i=1}^{n}
\sum_{\ell=1}^{\nu_j(x_{ij})}
\widehat{\Delta}_{j\ell}.
\label{eq:ale_effect}
\]
The second term centers the ALE function such that its empirical mean is zero.

In the real data implementation, each continuous predictor was divided into \(L=20\) quantile-based intervals. The ALE was calculated on the fitted Cox linear predictor scale; therefore, the vertical axis represents the centered contribution to the fitted log relative hazard. For graphical presentation, the curves are displayed from the first to the 99th empirical percentiles to reduce the visual influence of extreme observations.

\section{Numerical simulation}

This simulation study was designed to evaluate the predictive performance and interpretability of BRP-GABLE for different outcome types and signal structures. We evaluated both the rule-only and full specifications of BRP-GABLE and compared them with those of GABLE, RuleFit, an outcome-specific linear model, gradient boosting machines (GBM), and RF in settings involving binary, continuous, and time-to-event outcomes. To investigate both well-specified and mis-specified settings, we considered four data-generating scenarios: purely rule-based, purely linear, mixed linear and rule-based, and smooth nonlinear.

\subsection{Data-generating settings}

\noindent\textbf{Covariate generation and common settings.}

For all simulation settings, the predictor vector was generated as follows:
\(
\boldsymbol{x}=(x_1,\dots,x_{20})^\top,
\)
where
\(
x_j \stackrel{\mathrm{i.i.d.}}{\sim} N(0,1), \quad j=1,\dots,20.
\)
The sample size was set to \(N=1000\). For each simulated dataset, half of the observations were randomly assigned to the training set \(N_{train}\) and the other half to the test set \(N_{test}\). Within each simulation replication, the same training--test split was used for all the competing methods. Each simulation was repeated 100 times. Additional simulations were conducted with sample sizes of \(N=500\) and \(N=2000\) using the same fixed settings, and the corresponding results are presented in Appendix~\ref{app:sample_size}.

\noindent\textbf{True functions.}

The following four scenarios were considered:

\noindent\textbf{Scenario 1 (purely rule-based): }\(
\eta_1(\boldsymbol{x})=c_0\Bigl[
4\sum_{j=4}^{9}I(x_j>0)
+4I(x_4>0)I(x_5>0)
+4I(x_6>0)I(x_7>0)
-14
\Bigr].
\)

\noindent\textbf{Scenario 2 (purely linear): }\(
\eta_2(\boldsymbol{x})=x_1+x_2+x_3+x_4+x_5+x_6.
\)

\noindent\textbf{Scenario 3 (mixed linear and rule-based): }\(
\eta_3(\boldsymbol{x})=0.5(x_1+x_2+x_3+x_4+x_5+x_6)+0.5c_0
\Bigl[
4\sum_{j=4}^{9}I(x_j>0)
+4I(x_4>0)I(x_5>0)
+4I(x_6>0)I(x_7>0)
-14
\Bigr].
\)

\noindent\textbf{Scenario 4 (smooth nonlinear misspecified): }\(
\eta_4(\boldsymbol{x})
=
4\Bigl[
\frac{\sin(x_1)+1}{2}
+\log(1+|x_2|)
-\sqrt{|x_3x_4|}
\Bigr].
\)

Here, \(\eta_1(\boldsymbol{x})\) represents a purely rule-based signal, \(\eta_2(\boldsymbol{x})\) a purely linear signal, \(\eta_3(\boldsymbol{x})\) a mixed linear and rule-based signal, and \(\eta_4(\boldsymbol{x})\) a smooth nonlinear signal that is not explicitly specified for all of the methods. The constant \(c_0\), which was used in Scenarios 1 and 3, was selected by comparing the standard deviations of the linear and rule-based components:
\[
c_0=\frac{\mathrm{sd}(x_1+x_2+x_3+x_4+x_5+x_6)}{\mathrm{sd}\!\left(
4\sum_{j=4}^{9}I(x_j>0)
+4I(x_4>0)I(x_5>0)
+4I(x_6>0)I(x_7>0)
-14
\right)},
\]
which yields \(c_0\approx 0.36\). In Scenario 3, this scaling yielded linear and rule-based components with approximately balanced variance. The same scaling factor was applied to the purely rule-based signal in Scenario 1.

\noindent\textbf{Outcome generation.}

For the binary outcome setting, the response variable was generated from a Bernoulli distribution with success probability
\[
\pi(\boldsymbol{x})=\frac{\exp(\eta(\boldsymbol{x}))}{1+\exp(\eta(\boldsymbol{x}))},
\qquad
y\sim \mathrm{Bernoulli}\bigl(\pi(\boldsymbol{x})\bigr).
\]
For the continuous outcome setting, the response variable was generated as follows:
\[
y=\eta(\boldsymbol{x})+\varepsilon, \quad \varepsilon \sim N\!\left(0,1\right).
\]
For the time-to-event outcome setting, data were generated using the Cox proportional hazards model. Let \(t_i^\ast\) denote the true event time for the \(i\)-th observation, \(C_i\) the censoring time, \(t_i\) the observed time, and \(\delta_i\) the event indicator. The baseline hazard was fixed at \(\lambda_0=0.1\), and the individual hazard was defined as
\[
\lambda_i=\lambda_0\exp\bigl(\eta(\boldsymbol{x}_i)\bigr).
\]
The true event time was then generated as follows:
\[
t_i^\ast \sim \mathrm{Exponential}(\lambda_i).
\]
To introduce censoring, the censoring time was generated as follows:
\[
C_i \sim U(0,20).
\]
The observed time and event indicators were defined as follows:
\[
t_i=\min(t_i^\ast,C_i),\qquad
\delta_i=I(t_i^\ast \le C_i).
\]

\subsection{Compared methods and implementation settings}

We evaluated two specifications of BRP-GABLE. The rule-only specification uses only rules pooled across bootstrap samples, whereas the full specification combines pooled rules with linear terms. The competing methods used were GABLE, RuleFit, an outcome-specific linear model, GBM, and RF. The method labelled \textit{Linear} in Figures~\ref{fig:binary}--\ref{fig:survival} denotes logistic regression for binary outcomes, ordinary linear regression for continuous outcomes, and the Cox proportional hazards model for time-to-event outcomes.

The GBM was implemented using the \texttt{gbm} package (version 2.3.1) \cite{ridgeway2026gbm}. RF was implemented using the \texttt{randomForest} package (version 4.7-1.2) for binary and continuous outcomes \cite{liaw2002randomforest} and the \texttt{randomForestSRC} package (version 3.6.2) for time-to-event outcomes \cite{ishwaran2008randomsurvival,ishwaran2026randomforestsrc}. RuleFit was implemented using the \texttt{pre} package (version 1.0.7) \cite{fokkema2020pre}. The lasso-based estimation for BRP-GABLE was performed using the \texttt{glmnet} package \cite{friedman2010}. These methods were fitted using the default settings of their respective packages. In the primary analysis, RuleFit used the \texttt{pre} package default of \texttt{maxdepth=3}, whereas the maximum rule degree in BRP-GABLE was set to \(d_{\max}=2\). To examine whether the maximum rule degree setting influenced the results, RuleFit was additionally fitted with \texttt{maxdepth=2}, while all other settings remained at their package defaults. The same simulated datasets, training--test splits, and random seeds as those in the primary analysis were used. The results of this matched-depth sensitivity analysis are presented in Appendix~\ref{app:rulefit_depth}.

For GABLE and each bootstrap-specific rule construction procedure in BRP-GABLE, the maximum number of generated rule terms was set to \(M_{\max}=10\) and the maximum interaction degree was set to \(d_{\max}=2\). A candidate cutoff was required to leave at least 5\% of the training observations for each child rule. Within the resulting admissible rank interval, the cutoff candidates were evaluated on a centered equally spaced rank grid with a spacing of 20\% of the training sample size. For the GABLE comparator, the forward-generated rule terms were subsequently reduced via stepwise selection using the \texttt{stepAIC} function in the \texttt{MASS} package~\cite{venables2002}. The penalty multiplier was fixed at \(k=6\), corresponding to the modified information criterion \(-2\log L(\eta)+2\xi q\) with \(\xi=3\) used in the time-to-event extension of ABLE by Shimokawa et al.~\cite{shimokawa2014}. The same fixed penalty value was applied to all three outcome types. For BRP-GABLE, the number of bootstrap replications was set to \(B=50\).

For both BRP-GABLE specifications, the penalty parameter for the final LASSO model was selected using 10-fold cross-validation and the one-standard-error rule. Rule-only and full specifications were fitted using the same bootstrap rule pool, training--test split, and cross-validation folds within each simulation replication.

\subsection{Evaluation criteria}

\noindent\textbf{Predictive performance.}
For binary outcomes, predictive performance was evaluated using the area under the receiver operating characteristic curve (AUC). For continuous outcomes, predictive performance was evaluated using the mean squared error (MSE). For time-to-event outcomes, predictive performance was evaluated using the concordance index (C-index).

\noindent\textbf{Interpretability.}
Interpretability was evaluated from the perspective of structural simplicity. We used two complementary measures: the total number of selected terms in the final model and the proportion of selected terms with a degree of one. These measures were calculated for GABLE, RuleFit, and the rule-only and full specifications of BRP-GABLE because the selected terms in all four model specifications admit a comparable rule-based representation.

The total term count is defined as the number of nonzero rules and linear terms retained in the final model, excluding the intercept. Thus, for models containing both rule and linear terms, the total term count includes both predictor types. A smaller total term count indicates a more concise model.

The degree of a rule is defined as the number of indicator conditions within the rule. For example,
\(
I(x_1>1)
\)
has degree 1, whereas
\(
I(x_2\leq 1)I(x_3>2)
\)
has degree 2. Each selected linear term is treated as a degree-1 term because it represents a main effect involving a single predictor.

Let \(N_{\mathrm{rule},1}\) denote the number of selected degree-1 rules, \(N_{\mathrm{linear}}\) the number of selected linear terms, and \(N_{\mathrm{total}}\) the total number of selected rules and linear terms. The proportion of degree-1 terms is defined as
\[
P_{\mathrm{degree}\text{-}1}
=
\frac{
N_{\mathrm{rule},1}+N_{\mathrm{linear}}
}{
N_{\mathrm{total}}
}.
\]
A larger value indicates that the selected model is more strongly concentrated on main effects and one-condition rules. This proportion is interpreted jointly with the total term count because a high degree-1 proportion alone does not necessarily imply that the overall model is small.

\subsection{Simulation results}

The simulation results are summarized in Figures~\ref{fig:binary}, \ref{fig:continuous}, and \ref{fig:survival}, which correspond to the binary, continuous, and time-to-event outcomes, respectively. For each outcome type, the four rows correspond to Scenarios 1--4, representing purely rule-based, purely linear, mixed linear and rule-based, and mis-specified smooth nonlinear settings, respectively. The first column reports the predictive performance measured by AUC, MSE, or C-index, according to the outcome type. The second and third columns report the final term count and the proportion of degree-1 terms, respectively.

\subsubsection{Binary outcome}

\begin{figure}[ht]
    \centering
    \includegraphics[width=1\linewidth]{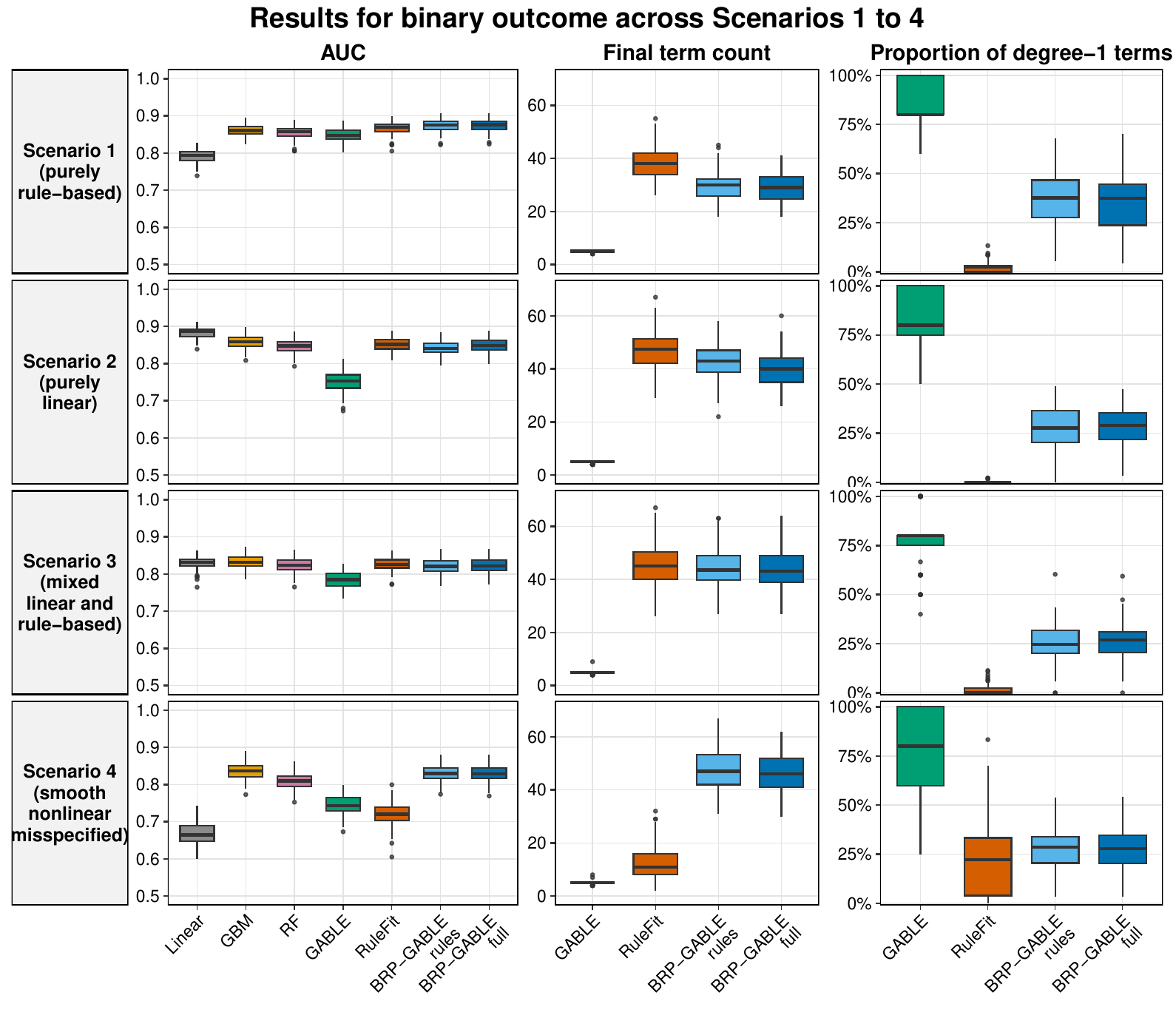}
    \caption{Results for the binary outcome across Scenarios 1--4. The four rows correspond to Scenarios 1--4, and the three columns show AUC, final term count, and proportion of degree-1 terms, respectively.}
    \label{fig:binary}
\end{figure}

\paragraph{Predictive accuracy.}
As shown in Figure~\ref{fig:binary}, both BRP-GABLE specifications consistently improved upon GABLE across the four scenarios. In Scenario~1, in which the true signal was purely rule-based, BRP-GABLE achieved the highest average AUC among the compared methods. In the purely linear Scenario~2, the linear model performed the best, as expected. Full BRP-GABLE improved upon its rule-only counterpart and performed comparably to RuleFit. In the mixed Scenario~3, BRP-GABLE remained competitive with the other flexible methods, although GBM and the linear model achieved slightly higher average AUCs. The results for Scenario~4 are particularly notable. Although the smooth nonlinear signal was not represented directly by the rule-generating model, both BRP-GABLE specifications substantially outperformed GABLE and RuleFit. They also outperformed RF and approached the predictive accuracy of GBM.

\paragraph{Interpretability.}
Relative to RuleFit, BRP-GABLE retained fewer final terms in Scenario 1 and a broadly comparable number in Scenarios 2 and 3. More importantly, the proportion of degree-1 terms was substantially higher in these structured scenarios. This indicates that even when the total model sizes were similar, BRP-GABLE represented the fitted relationship using a larger proportion of main-effect structures and fewer interaction terms. GABLE remained the sparsest method in terms of the final term count; however, this sparsity was accompanied by consistently lower predictive accuracy. Scenario~4 showed a different pattern in which BRP-GABLE retained more terms than the other baseline methods.

\subsubsection{Continuous outcome}
\begin{figure}[ht]
    \centering
    \includegraphics[width=1\linewidth]{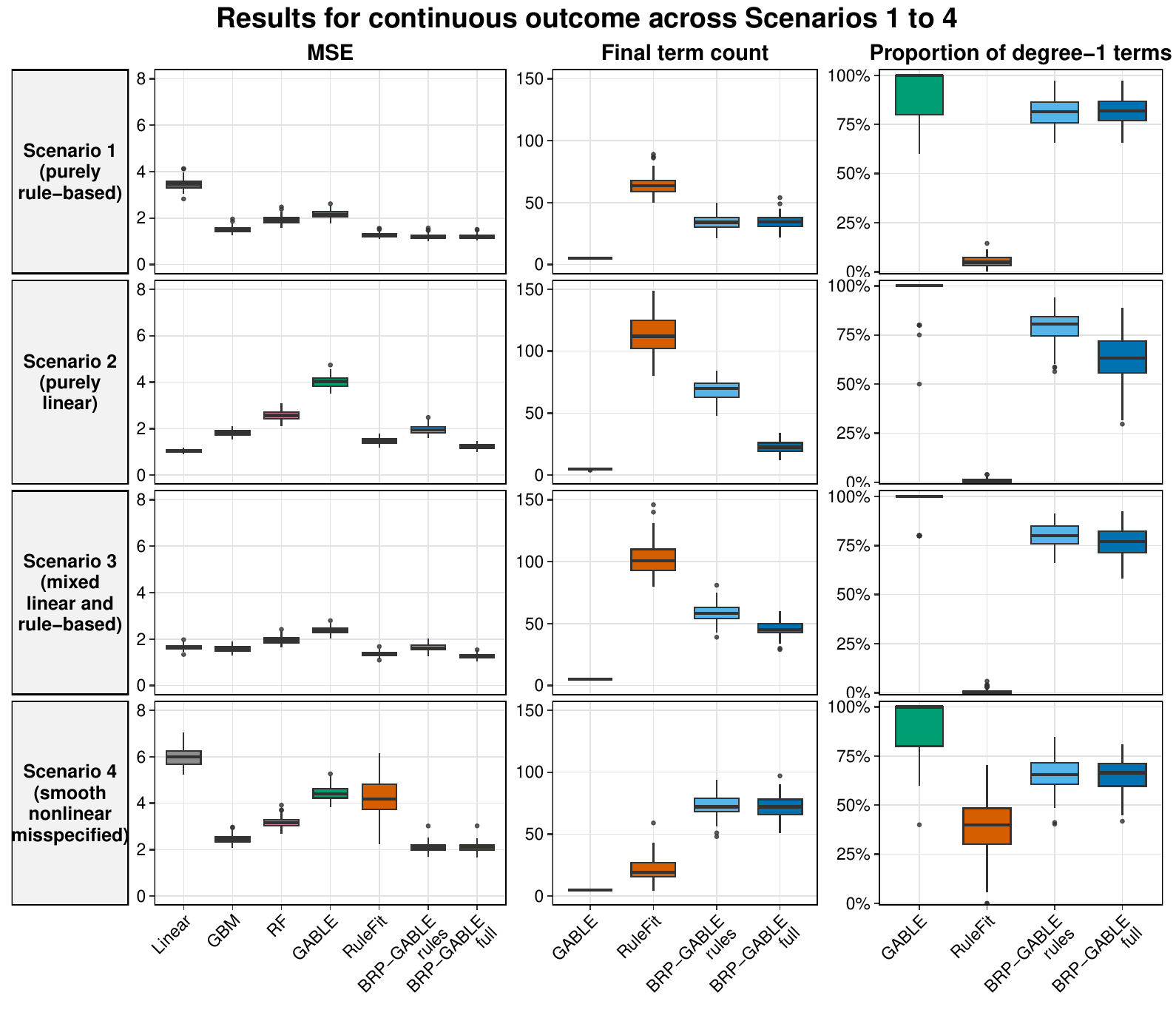}
    \caption{Results for the continuous outcome across Scenarios 1--4. The four rows correspond to Scenarios 1--4, and the three columns show MSE, final term count, and proportion of degree-1 terms, respectively.}
    \label{fig:continuous}
\end{figure}
\paragraph{Predictive accuracy.}
Figure~\ref{fig:continuous} shows a clear improvement in both BRP-GABLE specifications over GABLE for all four scenarios. BRP-GABLE achieved the lowest or nearly the lowest average MSE in the purely rule-based Scenario~1. In the purely linear Scenario~2, the linear model performed best, while full BRP-GABLE remained comparable and substantially outperformed rule-only BRP-GABLE. The advantage of the full specification was also apparent in the mixed Scenario~3, where BRP-GABLE achieved the lowest average MSE among all methods. In Scenario~4, both BRP-GABLE specifications outperformed not only GABLE and RuleFit but also GBM and RF.

\paragraph{Interpretability.}
In Scenarios~1--3, BRP-GABLE generally retained fewer final terms than RuleFit while maintaining a substantially higher proportion of degree-1 terms. The differences between the two BRP-GABLE specifications were particularly informative in purely linear and mixed settings. By directly representing the linear effects, full BRP-GABLE required markedly fewer final terms than the rule-only specification. In Scenario~4, BRP-GABLE retained considerably more final terms than RuleFit. Nevertheless, the selected models still had a higher proportion of degree-1 terms and achieved substantially better predictive accuracy.

\subsubsection{Time-to-event outcome}

\begin{figure}[ht]
    \centering
    \includegraphics[width=1\linewidth]{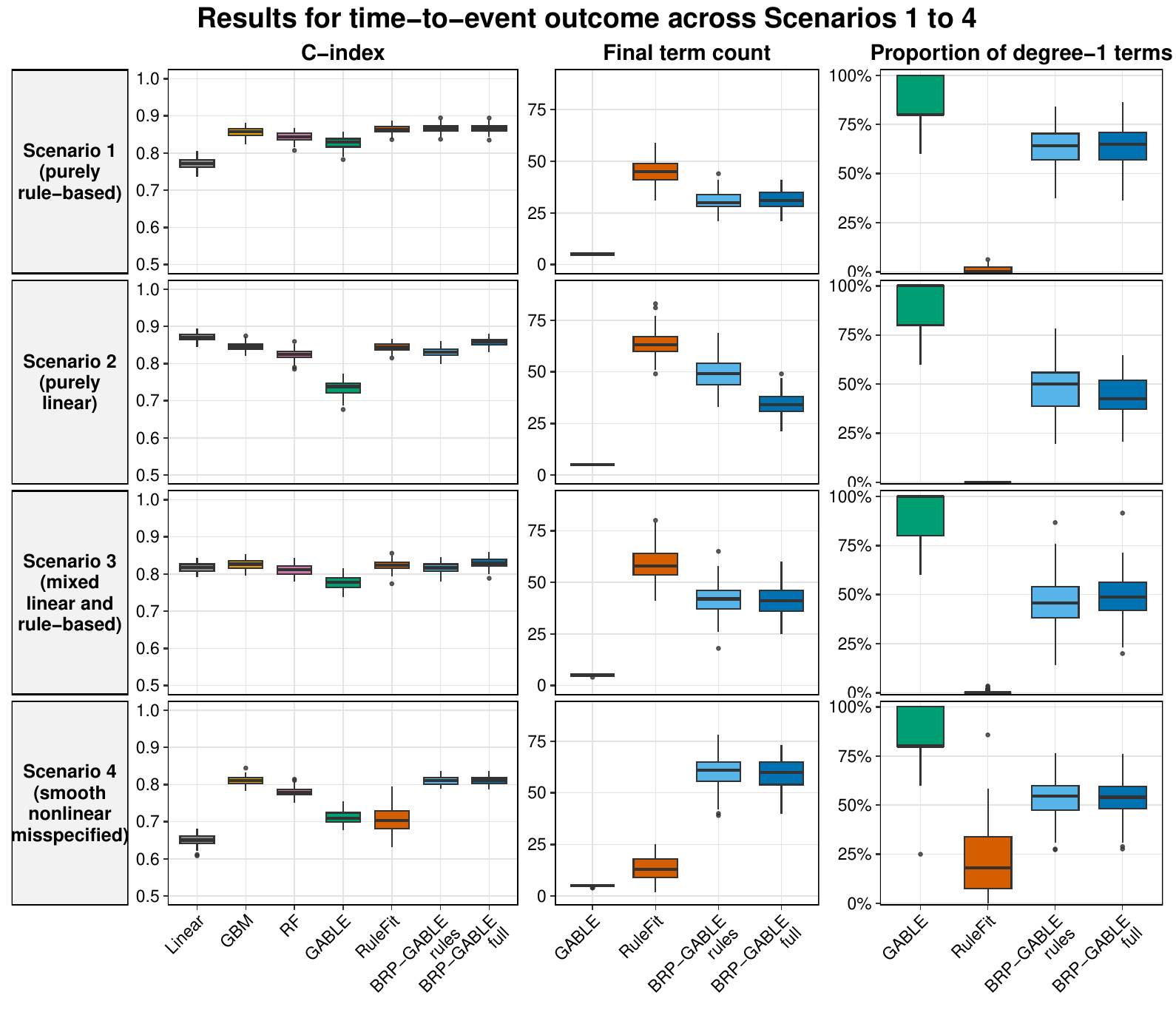}
    \caption{Results for the time-to-event outcome across Scenarios 1--4. The four rows correspond to Scenarios 1--4, and the three columns show C-index, final term count, and proportion of degree-1 terms, respectively.}
    \label{fig:survival}
\end{figure}

\paragraph{Predictive accuracy.}
For the time-to-event outcome, Figure~\ref{fig:survival} again shows that both BRP-GABLE specifications consistently outperformed GABLE. BRP-GABLE achieved the highest or nearly the highest average C-index in the purely rule-based Scenario~1. In the purely linear Scenario~2, the linear model performed best, while full BRP-GABLE outperformed the rule-only specification and all other nonlinear baseline methods. Full BRP-GABLE also achieved the highest average C-index in the mixed Scenario~3. In Scenario~4, its predictive performance was virtually identical to that of GBM and the rule-only specification, while clearly exceeding that of RF, RuleFit, and GABLE.

\paragraph{Interpretability.}
Across Scenarios~1--3, BRP-GABLE generally retained fewer final terms than RuleFit and consistently exhibited a markedly higher proportion of degree-1 terms. The effect of the linear component was particularly clear in Scenario~2, where full BRP-GABLE used fewer final terms than both RuleFit and rule-only BRP-GABLE, while simultaneously attaining better predictive performance. In Scenario~4, BRP-GABLE retained more final terms than RuleFit. However, the selected model contained a higher proportion of degree-1 terms and achieved substantially better predictive accuracy than RuleFit, while remaining comparable to GBM.

\paragraph{Overall summary.}
Taken together, the simulation results show that BRP-GABLE consistently improved the predictive performance of GABLE across all outcome types and scenarios. A comparison between rule-only BRP-GABLE and full BRP-GABLE illustrates the contribution of the explicit linear component. The two specifications performed similarly in settings dominated by rule-based or smooth nonlinear signals, whereas the full specification generally performed better and required fewer final terms when the true model contained linear effects.

The inclusion of GBM and RF further clarified the predictive role of BRP-GABLE. In the misspecified smooth nonlinear Scenario~4, BRP-GABLE outperformed both black-box benchmarks for the continuous outcome and performed comparably to the best black-box method for the binary and time-to-event outcomes.

Finally, the final term count and degree-1 proportion provide complementary evidence for interpretability. In Scenarios~1--3, BRP-GABLE often used fewer or a comparable number of terms relative to RuleFit while consistently assigning a much larger proportion of the fitted model to degree-1 terms. In Scenario~4, this advantage was accompanied by a larger final model, reflecting the additional flexibility required to approximate a misspecified smooth signal. Overall, BRP-GABLE provided a favorable balance between predictive performance, model size, and structural simplicity, although this balance depended on the underlying signal structure.

\section{Real-data implementation}

To further examine the practical usefulness of BRP-GABLE, we applied the proposed method to a real clinical time-to-event dataset. In this section, we focus on overall survival and investigated both the predictive performance and model interpretability in a real data setting. The analysis comprised two experiments. Experiment~1 compared the predictive performance and model complexity of BRP-GABLE with those of competing methods over repeated random train--test splits. Experiment~2 focused exclusively on the final BRP-GABLE model fitted to the full dataset and examined the fitted model from complementary perspectives, including its retained terms, rule activation patterns, variable-level effects, and time-to-event outcomes of rule-defined subgroups.

\subsection{MGUS data and implementation setting}

We used the \texttt{mgus2} dataset from the \texttt{survival} package~\cite{survival-package}, which is based on a long-term Mayo Clinic cohort of patients diagnosed with monoclonal gammopathy of undetermined significance (MGUS). The original clinical study investigated the risk of progression from MGUS to multiple myeloma and other related plasma cell disorders \cite{kyle2002}. However, in the present implementation, we used the overall survival endpoints available in the \texttt{mgus2} dataset. This choice was motivated by long-term follow-up evidence showing that patients with MGUS have a shorter overall survival than the age- and sex-matched general population \cite{kyle2018}.

The time-to-event outcome was defined as the follow-up time from MGUS diagnosis to death or last contact, measured in months, and the corresponding death indicator. Specifically, \texttt{futime} was used as the follow-up time and \texttt{death} was used as the event indicator. The baseline predictors included age at diagnosis (\texttt{age}), sex (\texttt{sex}) coded as \(\texttt{male=1}\) and \(\texttt{female=0}\), year of diagnosis (\texttt{dxyr}), hemoglobin level (\texttt{hgb}), serum creatinine level (\texttt{creat}), and the size of the monoclonal serum spike (\texttt{mspike}). Patients with missing data for any of the selected variables were excluded. Specifically, 46 patients were excluded due to missing values, and the resulting complete-case dataset of 1,338 patients was used for the analysis. A total of 938 deaths were recorded among the 1338 patients. All methods were implemented using the same settings as those used in the simulation study.

\noindent\textbf{Experiment 1: empirical comparison of predictive performance.}
For each repetition, the complete-case dataset was randomly divided into training and test sets in a ratio of 7:3. The same random split was used for all the compared methods, and the procedure was repeated 100 times. The methods compared were the linear Cox model, GBM, RF, GABLE, RuleFit, rule-only BRP-GABLE, and full BRP-GABLE with both rule and linear terms. Predictive performance was evaluated on the test set using the C-index. For the four rule-based models---GABLE, RuleFit, rule-only BRP-GABLE, and full BRP-GABLE---the model complexity was additionally summarized by the final selected term count and the proportion of degree-1 terms among all retained terms. The selected linear terms were included in the final term count and treated as degree-1 terms. The linear Cox model, GBM, and RF were excluded from the complexity comparison, because they did not provide a directly comparable selected-rule decomposition.

\noindent\textbf{Experiment 2: explanation of the final fitted model.}
To demonstrate how BRP-GABLE can be used for model interpretation in real data applications, the proposed method was fitted once to the full complete-case MGUS2 dataset. Unlike Experiment~1, the purpose of this analysis was not to compare predictive performance across methods, but to examine the fitted BRP-GABLE model from multiple complementary perspectives.

First, all retained linear and rule terms are listed together with their coefficients, hazard ratios, and scale-adjusted importance values. A rule-support-effect map is then used to summarize multiple characteristics of the selected rules in a single display. In this map, the horizontal axis represents the empirical support for each rule, whereas the vertical axis represents the model-based hazard ratio associated with rule activation. Point size indicates relative importance, and point color indicates the rule degree. This display allows rule support, model-based associations, relative importance, and structural complexity to be jointly assessed. At the individual level, a patient--rule activation heatmap was used to display the selected rules satisfied by each patient. The pairwise overlap between patient sets satisfying different selected rules was descriptively summarized using the Jaccard similarity coefficient \cite{jaccard1901}, which ranges from 0 for no overlap to 1 for identical patient sets. At the variable level, the first-order accumulated local effect (ALE) curves were calculated following Apley and Zhu \cite{apley2020}, as defined in Section~2.4. Finally, Kaplan--Meier curves \cite{kaplan1958} were used to display the overall survival distributions of patients satisfying and not satisfying the representative rules selected by the final model. These curves were used as descriptive summaries of rule-defined patient subgroups and were not interpreted as estimates of causal effects.

\subsection{Results}

\begin{figure}[htbp]
\centering
\includegraphics[width=1\linewidth]{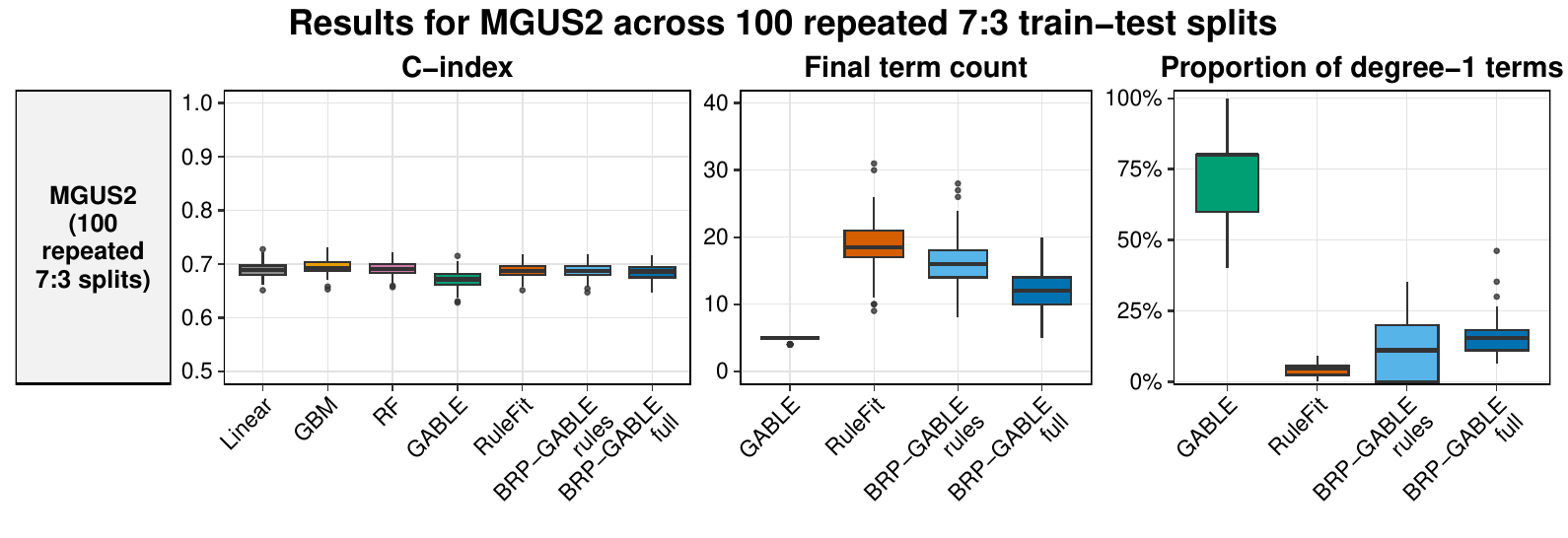}
\caption{Results on the MGUS2 dataset over 100 repeated random 7:3 train--test splits. The three columns show C-index, final term count, and proportion of degree-1 terms, respectively.}
\label{fig:mgus2_empirical}
\end{figure}

\noindent\textbf{Experiment 1: empirical comparison of predictive performance.}
Figure~\ref{fig:mgus2_empirical} summarizes the results of the seven methods for 100 repeated random 7:3 train--test splits of the MGUS2 dataset. In terms of predictive performance, GBM showed the highest mean C-index, followed by RF. The C-index distributions of the linear model, RuleFit, rule-only BRP-GABLE, and full BRP-GABLE overlapped. Rule-only BRP-GABLE showed a slightly higher C-index than full BRP-GABLE, whereas both specifications substantially outperformed GABLE. Overall, BRP-GABLE achieved predictive performance comparable to RuleFit and the other baseline methods.

Regarding model complexity, GABLE retained the fewest final terms, with full BRP-GABLE retaining the next fewest. Full BRP-GABLE retained fewer final terms than both rule-only BRP-GABLE and RuleFit, whereas RuleFit had the largest final term count among the four rule-based models. The proportion of degree-1 terms was highest for GABLE. Among the models with comparable predictive performance, full BRP-GABLE had a higher proportion of degree-1 terms than rule-only BRP-GABLE and RuleFit. Thus, relative to RuleFit, full BRP-GABLE showed a smaller final term count and a larger proportion of degree-1 terms, while maintaining a broadly similar C-index distribution.

\noindent\textbf{Experiment 2: multifaceted explanation of the final fitted model.}
Table~\ref{tab:rule_importance_mgus2} lists all the terms retained in the final model. Identifiers R1--R7 were assigned to the selected rules in decreasing order of relative importance and are used consistently in subsequent figures. Here, \textit{Coef.} denotes the coefficient transformed back to the original scale and \textit{HR} denotes the corresponding hazard ratio. For the linear age term, the hazard ratio represents the change associated with a one-year increase in age. For a rule term, the hazard ratio compares activation and non-activation of the rule, while holding the remaining model terms fixed. The importance values were calculated using the scale-adjusted importance measure defined in Section~2.4, and rescaled so that the largest value among all retained terms was 100.

\begin{table}[htbp]
\centering
\small
\renewcommand{\arraystretch}{1.2}
\setlength{\tabcolsep}{5pt}
\caption{Terms retained in the final BRP-GABLE model fitted to the complete-case MGUS2 dataset}
\label{tab:rule_importance_mgus2}
\begin{tabular}{clccc}
\toprule
ID & Retained term & Coef. & HR & Rel. importance \\
\midrule
L1 & \texttt{age} (linear)                              &  0.020 & 1.020 & 100.00 \\
R1 & \texttt{age} $\leq 73$ AND \texttt{hgb} $> 11.7$   & -0.155 & 0.856 &  34.38 \\
R2 & \texttt{age} $\leq 75$ AND \texttt{creat} $\leq 1.2$ & -0.090 & 0.914 &  19.92 \\
R3 & \texttt{age} $\leq 72$ AND \texttt{hgb} $> 11.1$   & -0.059 & 0.943 &  13.07 \\
R4 & \texttt{age} $\leq 79$ AND \texttt{hgb} $> 12.6$   & -0.053 & 0.948 &  11.84 \\
R5 & \texttt{age} $\leq 78$ AND \texttt{creat} $\leq 1.7$ & -0.050 & 0.951 &  10.49 \\
R6 & \texttt{age} $\leq 72$ AND \texttt{hgb} $> 10.9$   & -0.023 & 0.977 &   5.05 \\
R7 & \texttt{age} $\leq 72$ AND \texttt{creat} $\leq 1.5$ & -0.002 & 0.998 &   0.52 \\
\bottomrule
\end{tabular}
\end{table}

The final model retained one linear term and seven degree-2 rules. The linear effect of age had the greatest relative importance. Among the selected rules, R1 had the largest importance and smallest hazard ratio, followed by R2 and R3. Hazard ratios for the seven selected rules ranged from 0.856 to 0.998. Their support, model-based effects, and relative importance were examined jointly in the following rule support--effect map:

\begin{figure}[htbp]\centering\includegraphics[width=0.7\linewidth]{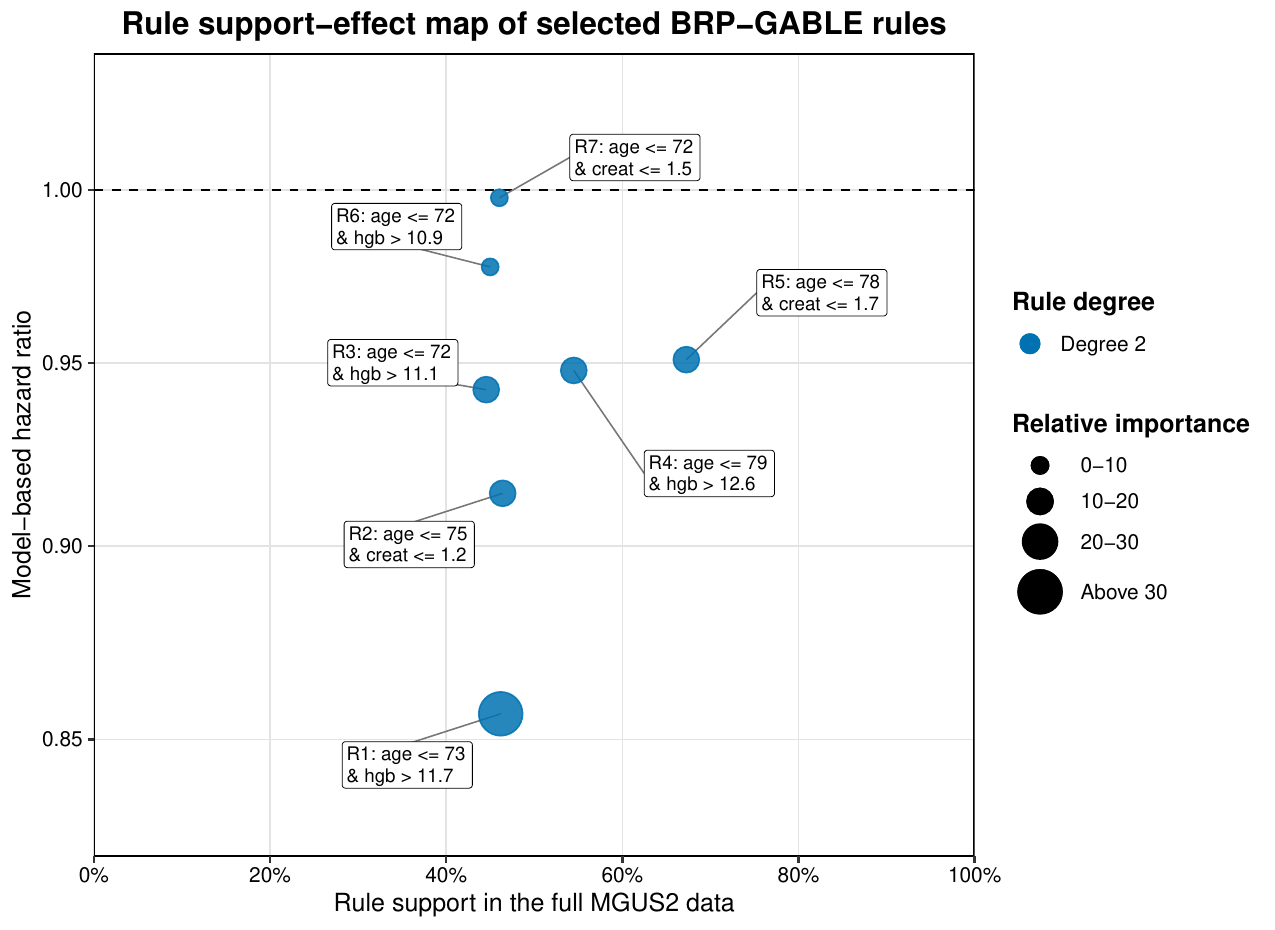}\caption{Rule support--effect map for the seven rules retained in the final BRP-GABLE model fitted to the full complete-case MGUS2 dataset.}
\label{fig:rule_support}
\end{figure}

Figure~\ref{fig:rule_support} summarizes the support, model-based effects, and relative importance of the selected rules. The horizontal axis represents the proportion of patients satisfying each rule and the vertical axis represents the model-based hazard ratio associated with rule activation. Point size represents the relative importance category. Point color represents the degree of the rule. The dashed horizontal line indicates a hazard ratio of 1. Rule support ranged from 44.5\% to 67.3\%, indicating that none of the retained rules represented a very small subgroup of patients. All seven rules had hazard ratios below or close to 1. R1 had the smallest hazard ratio and largest relative importance. R5 had the largest support, capturing approximately two-thirds of the complete-case dataset, but had a smaller model-based effect and relative importance than R1. In contrast, R7 had a hazard ratio close to one and the smallest relative importance among the selected rules. All retained rules were degree-2 rules involving age together with either hemoglobin or serum creatinine. Next, we evaluated how the seven retained rules were activated across individual patients and how strongly their rule-defined patient subgroups overlapped.

\begin{figure}[htbp]\centering\includegraphics[width=1.0\linewidth]{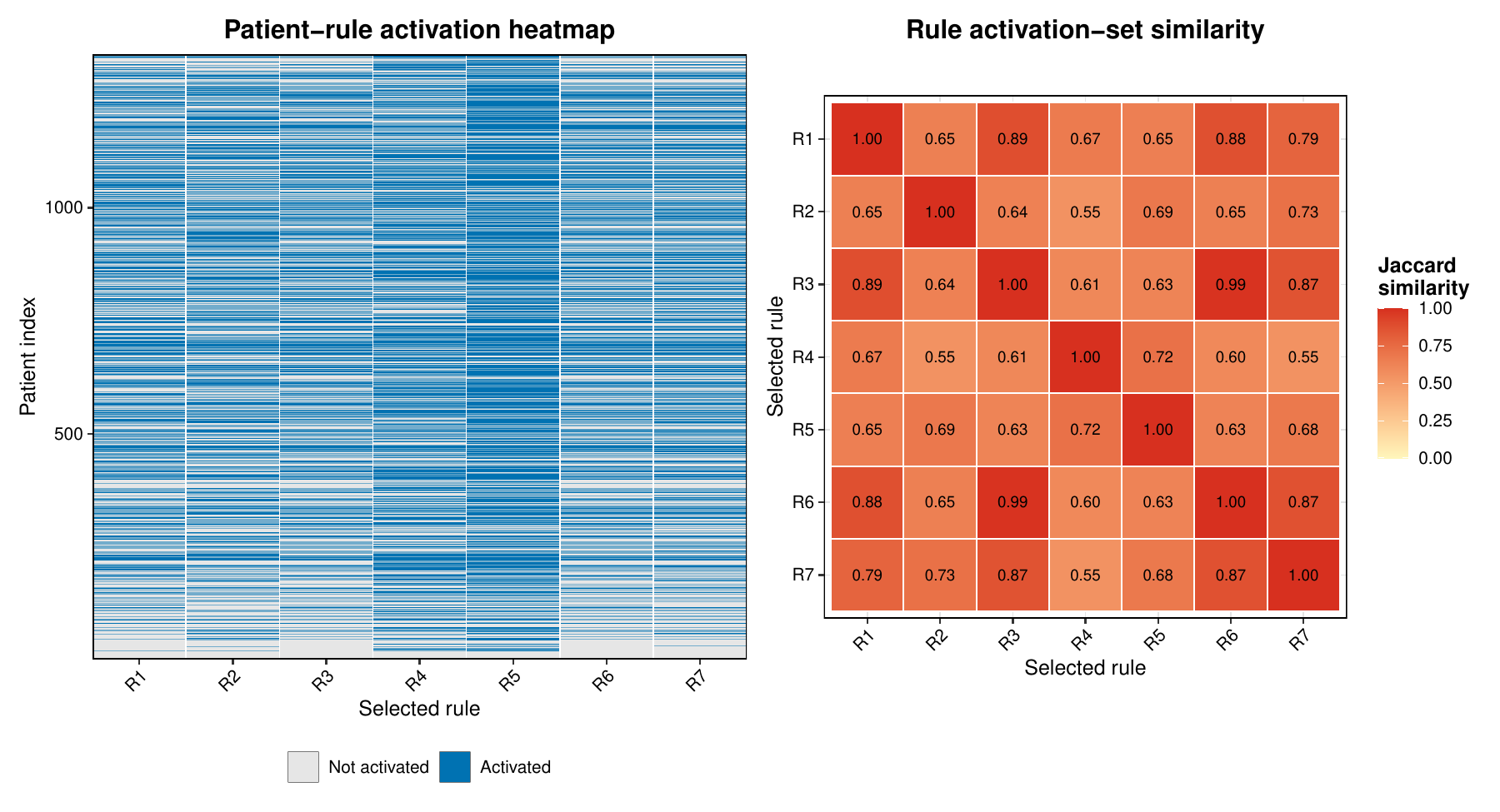}\caption{Patient-level activation patterns and pairwise activation-set similarities of the seven rules retained in the final BRP-GABLE model.}
\label{fig:heatmap}
\end{figure}

Figure~\ref{fig:heatmap} shows the activation structure of the seven retained rules at both the patient and rule levels. In the left panel, each row represents one patient, and each column represents one selected rule; blue cells indicate that the corresponding rule is satisfied, whereas gray cells indicate that it is not satisfied. The right panel shows the pairwise Jaccard similarities between patient sets that satisfied the selected rules. The rules are ordered from R1 to R7 according to their decreasing relative importance. The patient--rule activation heatmap shows that individual patients could satisfy multiple selected rules simultaneously and that the activation patterns differed across the seven rules. The pairwise Jaccard similarities ranged from 0.55 to 0.99, indicating that the rule-defined patient subgroups overlapped, rather than being mutually exclusive.

The highest similarity was observed between R3 and R6 (J=0.99). These two rules shared the same age condition and differed only slightly in their hemoglobin cutoffs. R1 also showed high similarity to R3 (J=0.89) and R6 (J=0.88). In contrast, R4 had the lowest similarity with R2 and R7 (J=0.55 for both pairs). Thus, although the selected rules represented distinct combinations of cutoff conditions, several captured substantially overlapping patient subgroups. Next, we examined the combined variable-level effects represented by the retained linear and rule terms using the accumulated local effect curves.

\begin{figure}[htbp]\centering\includegraphics[width=1.0\linewidth]{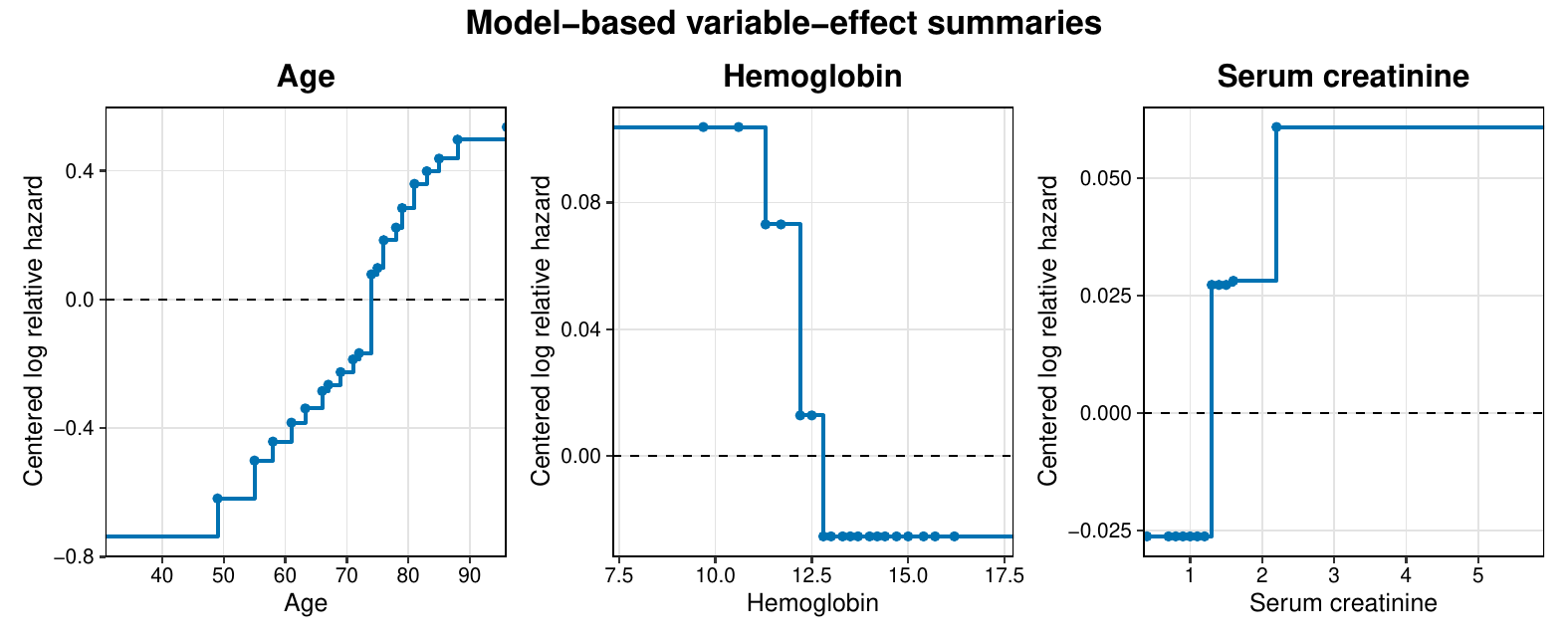}\caption{Accumulated local effects of age, hemoglobin, and serum creatinine in the final BRP-GABLE model fitted to the full complete-case MGUS2 dataset.}
\label{fig:ALE}
\end{figure}

Figure~\ref{fig:ALE} illustrates the accumulated local effects of three variables in the final model. The vertical axis represents the centered change in the fitted log relative hazard, with zero corresponding to the average reference level. The horizontal ranges were restricted to the first to 99th percentiles. Each curve summarizes the combined contributions of all retained linear and rule terms involving the corresponding variable. Among these, age exhibited the largest variation in the fitted log relative hazard. The age effect increased across the displayed range, with negative centered effects at younger ages and positive effects at older ages. The accumulated local effect of hemoglobin decreased in a stepwise manner as hemoglobin increased. Lower hemoglobin values were associated with positive centered effects, whereas higher values were associated with negative centered effects. Stepwise changes corresponded to different hemoglobin cutoffs represented in R1, R3, R4, and R6. By contrast, the accumulated local effects of serum creatinine levels increased across the range displayed. Lower creatinine values had negatively centered effects, followed by increases over the intermediate range and a positive plateau at higher values. This pattern summarizes the combined effects of the creatinine conditions in R2, R5, and R7.

Taken together, the ALE curves showed that across the observed predictor ranges, older age and higher serum creatinine levels were associated with a higher model-predicted hazard of death, whereas higher hemoglobin levels were associated with a lower model-predicted hazard. A supplementary qualitative comparison of the corresponding ALE curves obtained using RuleFit, RF, and GBM is provided in Appendix~\ref{app:ale_comparison}. Finally, we examined the overall survival distributions of the patient subgroups defined by the three most important rules using Kaplan--Meier curves.

\begin{figure}[htbp]\centering\includegraphics[width=1.0\linewidth]{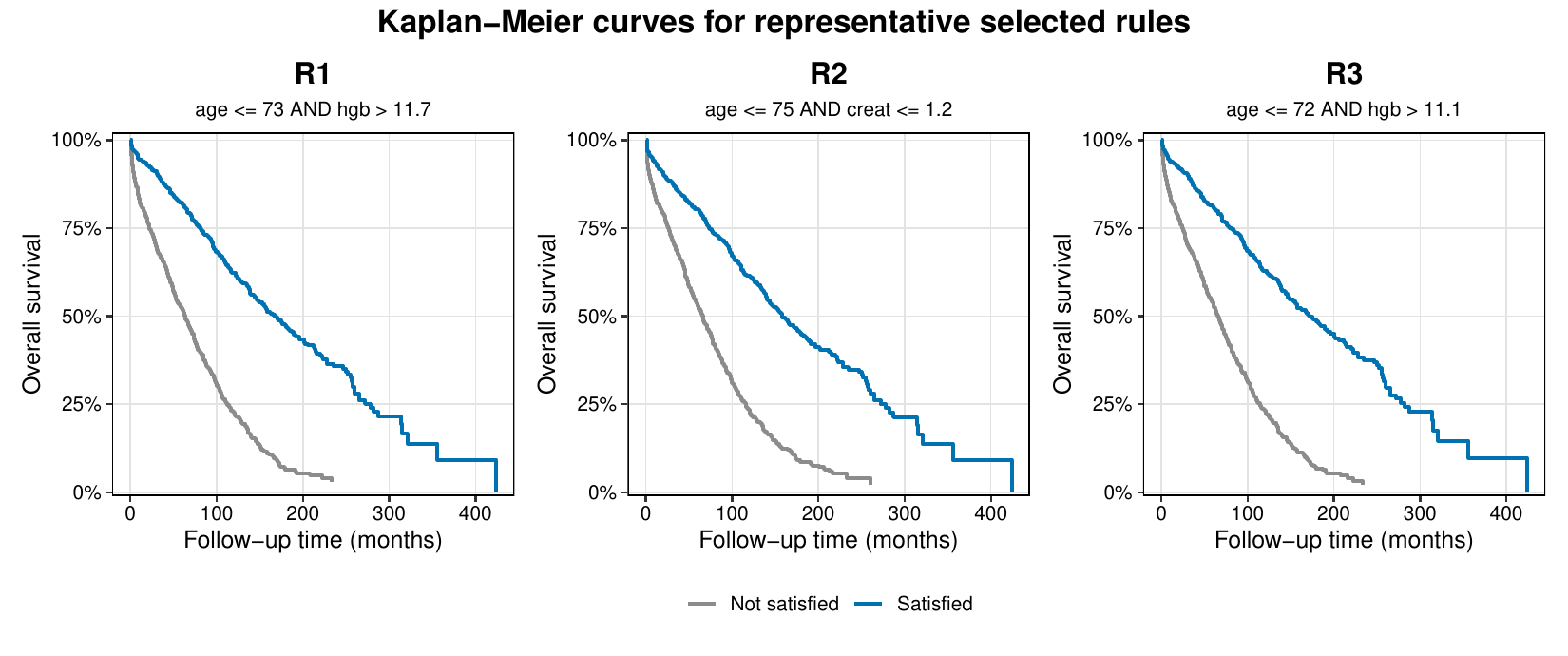}\caption{Kaplan--Meier curves for overall survival according to whether patients satisfied the three selected rules with the largest relative importance.}
\label{fig:kmcurve}
\end{figure}

Figure~\ref{fig:kmcurve} presents the Kaplan--Meier curves for the patient subgroups defined by R1, R2, and R3, which are the three selected rules with the largest relative importance. The blue and gray curves represent patients who satisfied and did not satisfy each rule, respectively. For all three rules, patients who satisfied the rules had higher estimated survival probabilities over most of the follow-up period than those who did not. This pattern was directionally consistent with the model-based hazard ratios below one reported for R1, R2, and R3.

\section{Discussion and conclusion}

BRP-GABLE was developed to broaden the rule search of GABLE, while retaining a single sparse and explicitly interpretable final model. Across the three outcome types, both specifications of BRP-GABLE consistently improved predictive performance relative to GABLE. This improvement is consistent with the intended role of bootstrap rule generation. Because a single GABLE fit follows a greedy forward path, its later candidate rules are constrained by earlier selection. Repeating this construction on different bootstrap samples makes alternative variables, cutoff values, and rule structures available for the final selection. Pooling these rules and jointly estimating their coefficients by LASSO therefore separates the exploration of candidate structures from the estimation of the final model. Importantly, the fitted bootstrap-specific models were not averaged and the bootstrap samples were used to expand the set of candidate rules considered by the final sparse model.

Overall, the simulations suggest that the relative performance of BRP-GABLE depends on the underlying signal structure. This method is particularly effective for threshold-based and mixed rule-linear signals, whereas the explicit linear component improves the representation of global linear effects. In smooth nonlinear settings, strong predictive performance was achieved at the cost of larger models, reflecting the need to approximate smooth functions with multiple piecewise-constant rules. Relative to RuleFit, BRP-GABLE generally favored lower-order structures, which is consistent with GABLE's ability to construct first-order threshold rules directly rather than inheriting conditions along tree paths. This structural advantage became more apparent with an increasing sample size and persisted in the matched-depth sensitivity analysis. In summary, BRP-GABLE offers a flexible compromise between predictive performance, sparsity, and low-order rule structure across a range of signal settings.

The MGUS2 analysis further showed that BRP-GABLE provided competitive survival prediction while retaining an interpretable model. The fitted model associated older age, lower hemoglobin, and higher serum creatinine levels with higher predicted mortality risk. The directions of the associations for age and serum creatinine were consistent with previous MGUS cohort studies reporting greater excess mortality among older patients and poorer overall survival among patients with elevated serum creatinine \cite{kristinsson2009,steiner2018}. These findings provide external clinical context for the fitted patterns, although they do not establish MGUS-specific or causal effects. Patient-level analyses also showed that distinctly selected rules could identify overlapping subgroups, reflecting the fact that BRP-GABLE removed only exact duplicate rules. The broadly similar ALE patterns obtained from RuleFit, GBM, and RF further indicated that the principal variable-level associations were not specific to the modelling structure of BRP-GABLE and provided additional support for the consistency of the fitted interpretation across modelling approaches. Together, the variable- and patient-level analyses illustrate how the fitted model can be examined from complementary perspectives.

One limitation is that the computational burden of BRP-GABLE was not formally evaluated in the present study. Future work should characterize how runtime and memory requirements scale with the sample size, number of predictors, number of bootstrap replications, and cutoff grid, and investigate more efficient implementations for larger datasets.

In conclusion, BRP-GABLE expands the rule structures considered by GABLE and performs a single global sparse selection over the resulting rule pool and linear terms. The simulation and MGUS2 results showed that it can provide competitive predictions across multiple outcome types while retaining an explicit model whose effects can be examined at the term, variable, and patient subgroup levels. BRP-GABLE therefore offers a unified and interpretable framework that combines flexible rule discovery, sparse global selection, and multilevel model interpretation for binary, continuous, and time-to-event outcomes.

\section*{Data and code availability}

The \texttt{mgus2} dataset is publicly available in the R
\texttt{survival} package. The code used for the simulation studies,
supplementary analyses, and real-data application is publicly available at
\url{https://github.com/Karasu259/BRP-GABLE}.

\newpage

\bibliographystyle{plain}
\bibliography{references}
\newpage
\appendix
\section{Additional simulation results for different sample sizes}
\label{app:sample_size}

To examine whether the simulation results depended on the sample size, we repeated the simulation experiments with total sample sizes of \(N=500\) and \(N=2000\), in addition to the main setting of \(N=1000\). Only the sample size was changed, and the data-generating mechanisms, model specifications, hyperparameter settings, train--test splitting procedure, and evaluation criteria remained unchanged. For each sample size, the generated data were randomly and equally divided into training and test sets. Thus, the training and test sample sizes were 250, 500, and 1000. Each combination of outcome type, scenario, and sample size was evaluated using 100 simulation replications. The same fixed-method configuration was applied to all three sample sizes. Specifically, the vertical axis range of the prediction accuracy panel was adapted for each scenario to facilitate comparison among methods.  Therefore, vertical scales should not be compared directly across scenarios.

\begin{figure}[htbp]
\centering
\includegraphics[width=\linewidth]
{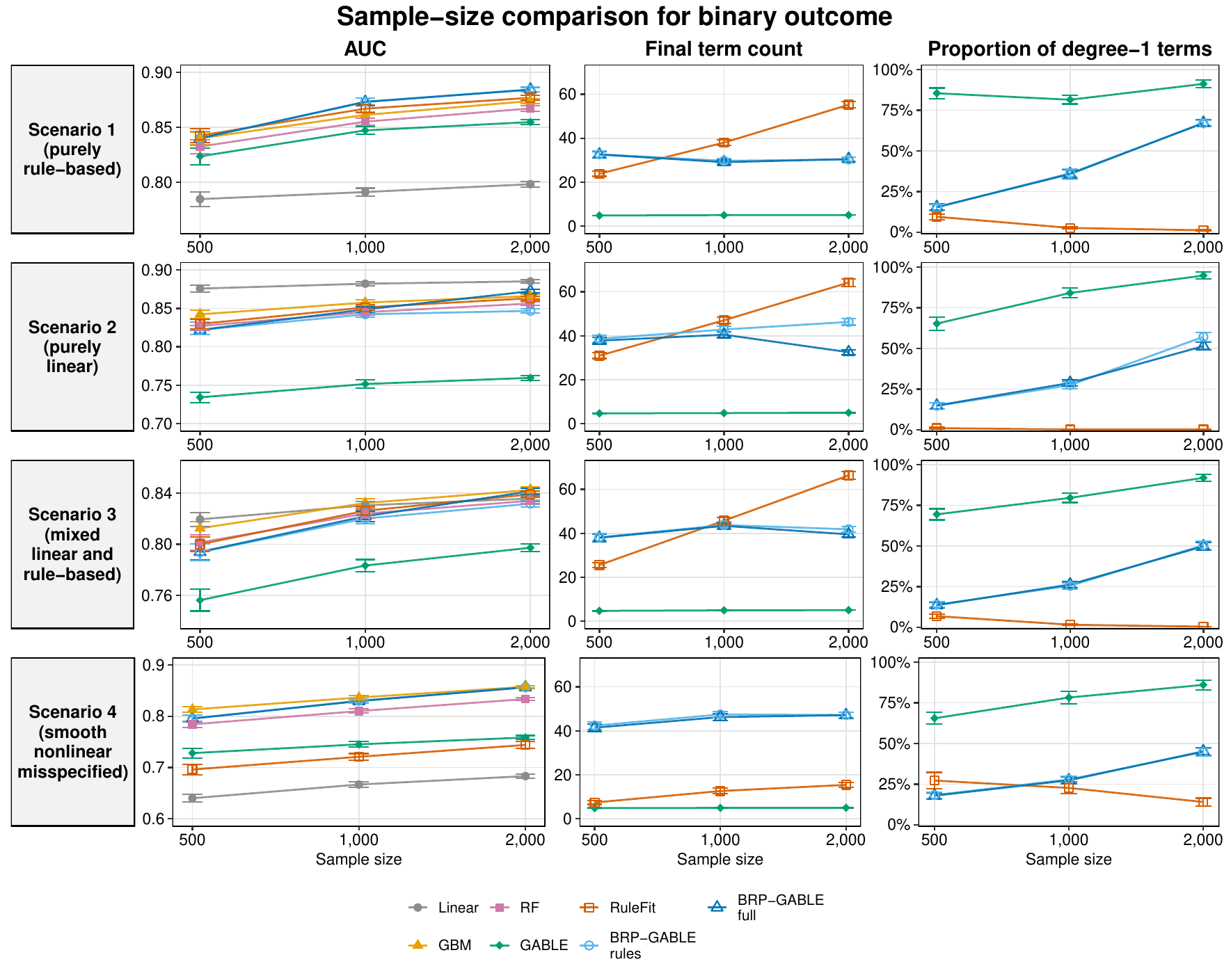}
\caption{Simulation results for the binary outcome across sample sizes \(N=500\), \(N=1000\), and \(N=2000\).}
\label{fig:sample_size_binary}
\end{figure}

\begin{figure}[htbp]
\centering
\includegraphics[width=\linewidth]
{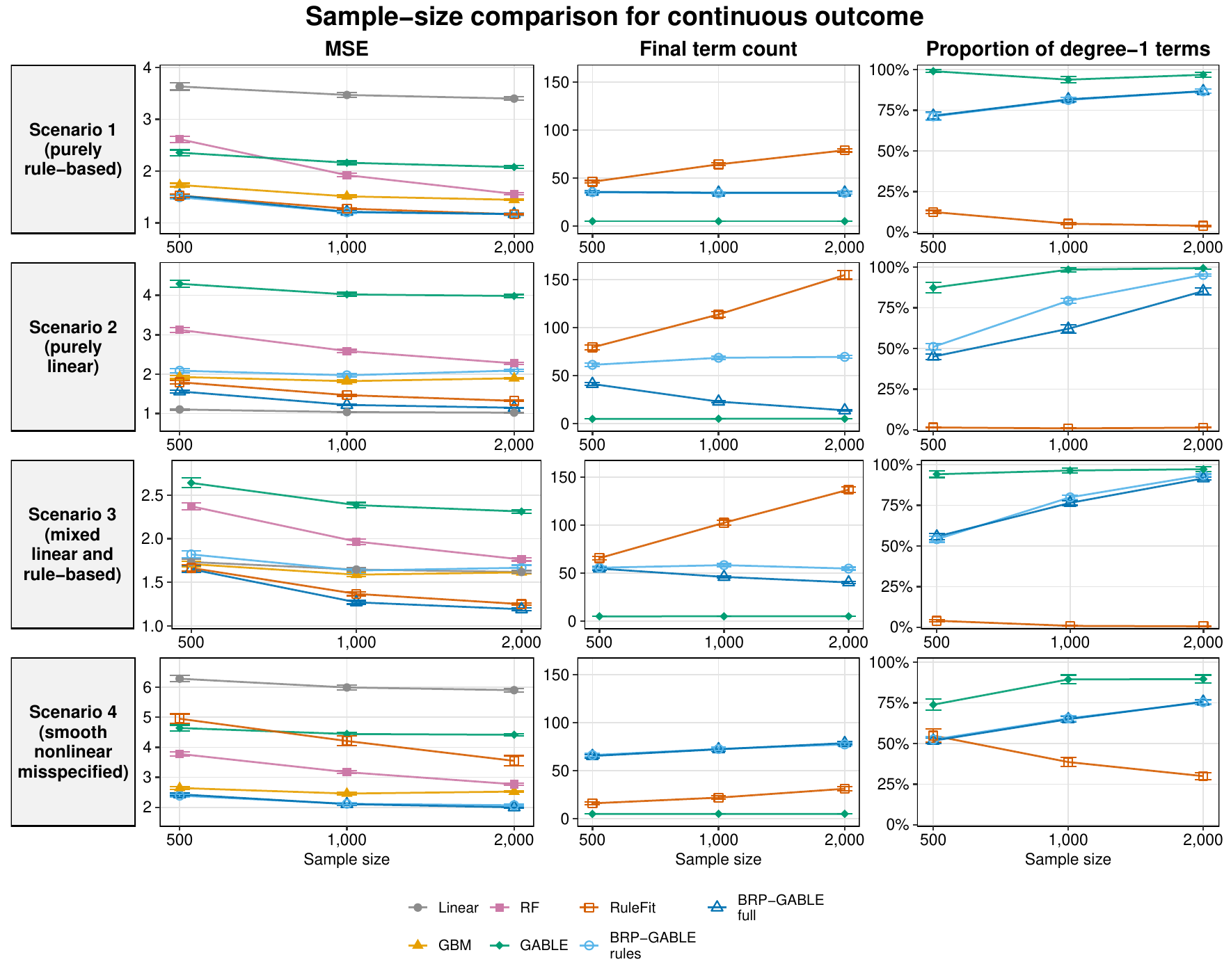}
\caption{Simulation results for the continuous outcome across sample sizes \(N=500\), \(N=1000\), and \(N=2000\).}
\label{fig:sample_size_continuous}
\end{figure}

\begin{figure}[htbp]
\centering
\includegraphics[width=\linewidth]
{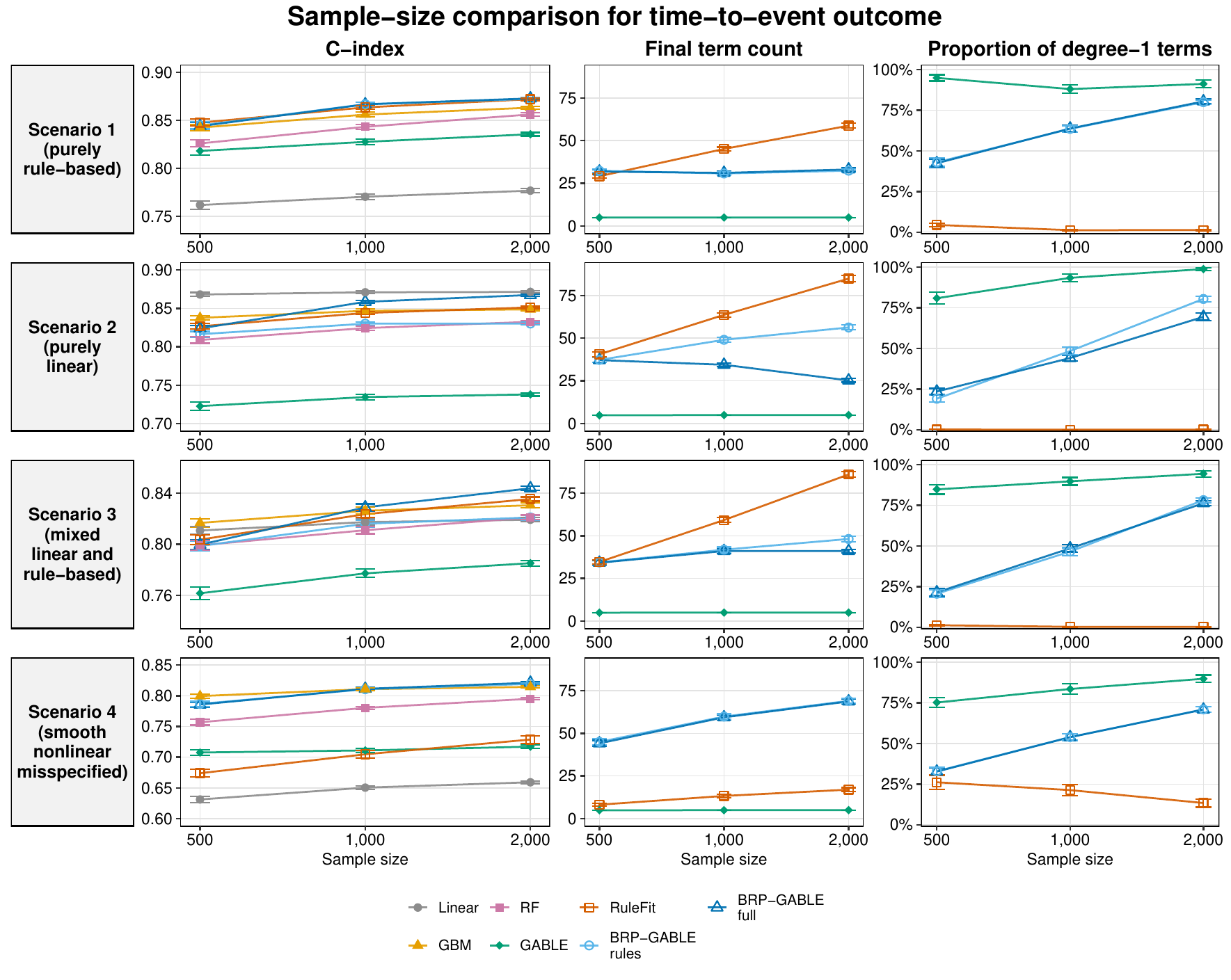}
\caption{Simulation results for the time-to-event outcome across sample sizes \(N=500\), \(N=1000\), and \(N=2000\).}
\label{fig:sample_size_time_to_event}
\end{figure}

Figures~\ref{fig:sample_size_binary}--\ref{fig:sample_size_time_to_event} compare the results for \(N=500\), \(N=1000\), and \(N=2000\). The plotted points represent the mean across 100 simulation replications, and the error bars represent the corresponding 95\% Monte Carlo confidence intervals for the mean. Predictive performance was reported for all methods, whereas the final term count and proportion of degree-1 terms were reported only for rule-based methods.

Across the three outcome types, predictive performance generally improved as the sample size increased. The AUC and C-index increased, whereas the MSE decreased. The relative predictive patterns of the compared methods remained broadly similar across the three sample sizes. Both BRP-GABLE specifications consistently achieved better predictive performance than GABLE. In Scenario~4, BRP-GABLE maintained a more favorable predictive performance than RuleFit and remained competitive with the black-box methods across the examined sample sizes.

The final term counts showed a clear dependence on the sample size. In Scenarios~1--3, the number of terms retained by RuleFit generally increased substantially with the sample size. The changes in the BRP-GABLE term counts were less pronounced in many of these structured settings, although increases were also observed in some settings, particularly in Scenario~4. The proportion of degree-1 terms retained by BRP-GABLE generally increased with sample size. At \(N=1000\) and \(N=2000\), both BRP-GABLE specifications retained substantially higher proportions of degree-1 terms than RuleFit across the three outcome types.

Overall, the principal predictive and structural patterns observed under the main setting of \(N=1000\) were also present at \(N=500\) and \(N=2000\). In particular, BRP-GABLE maintained competitive predictive performance while retaining a substantially larger proportion of degree-1 terms than RuleFit, with these structural differences becoming more apparent for larger sample sizes.

\FloatBarrier

\section{Supplementary comparison of accumulated local effects}
\label{app:ale_comparison}

As a supplementary qualitative comparison, first-order ALE curves were also calculated for RuleFit, GBM, and random survival forest (RF) fitted to the same full complete-case MGUS2 dataset. The comparison included all five continuous predictors: age at diagnosis (\texttt{age}), year of diagnosis (\texttt{dxyr}), hemoglobin level (\texttt{hgb}), serum creatinine level (\texttt{creat}), and the size of the monoclonal serum spike (\texttt{mspike}). For all four methods, the ALE was calculated using the same \(L=20\) quantile-based intervals and displayed from the first to the 99th empirical percentiles. As the prediction scores produced by the four methods were defined on different scales, each prediction function was divided by the standard deviation of its fitted prediction scores before the ALE was calculated. The resulting curves were centered and expressed on a standardized prediction score scale.

\begin{figure}[htbp]
\centering
\includegraphics[width=\linewidth]
{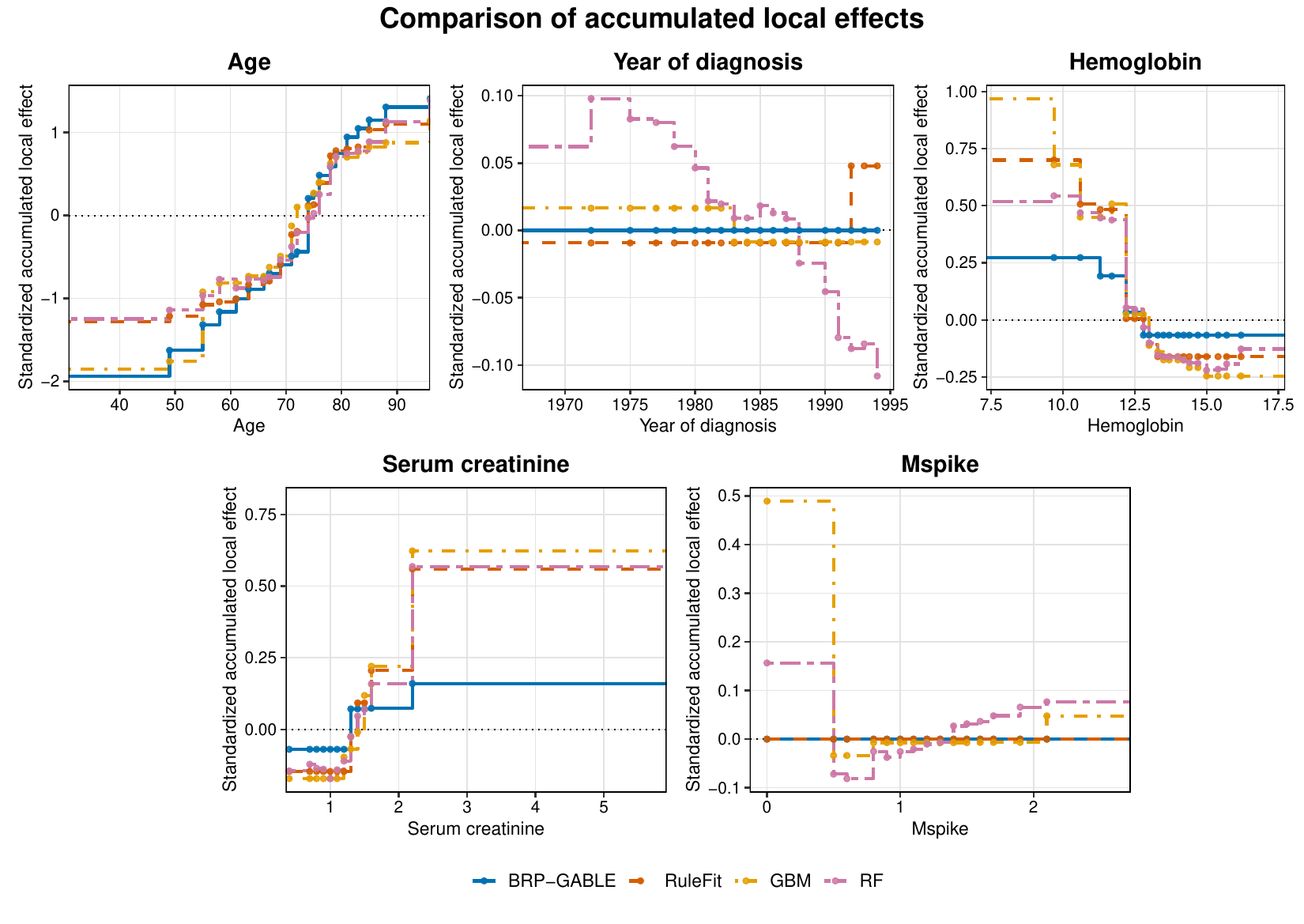}
\caption{Supplementary comparison of standardized first-order accumulated local effects from BRP-GABLE, RuleFit, GBM, and RF fitted to the full complete-case MGUS2 dataset.}
\label{fig:appendix_ale_comparison}
\end{figure}

Figure~\ref{fig:appendix_ale_comparison} shows that all four methods produced consistent effect directions for the three continuous variables represented in the final BRP-GABLE model. The standardized predicted risk generally increased with age, decreased with hemoglobin, and increased with serum creatinine. Although the magnitudes and detailed step structures differed among the methods, the principal variable-level patterns were qualitatively consistent. For the year of diagnosis and the size of the monoclonal serum spike, the curves remained close to zero over most of the observed ranges, with only small or localized deviations for some methods. These deviations were not consistently reproduced across the models; therefore, the two predictors did not show a consistent substantial effect across methods.

\FloatBarrier

\section{Analysis with a maximum RuleFit depth of two}
\label{app:rulefit_depth}

In the primary simulation analysis, RuleFit was fitted using the default settings of the \texttt{pre} package, under which the rules could contain up to three conditions. In contrast, BRP-GABLE was restricted to rules with a maximum degree of two. To assess whether this difference affected the comparison, RuleFit was refitted using \texttt{maxdepth=2}. All other RuleFit settings were retained at their package defaults. The simulated datasets, train--test splits, and RuleFit random seeds were identical to those used in the primary analysis; only RuleFit was refitted. The existing results for the two BRP-GABLE specifications were reused for comparison.

\begin{figure}[htbp]
\centering
\includegraphics[width=\linewidth]{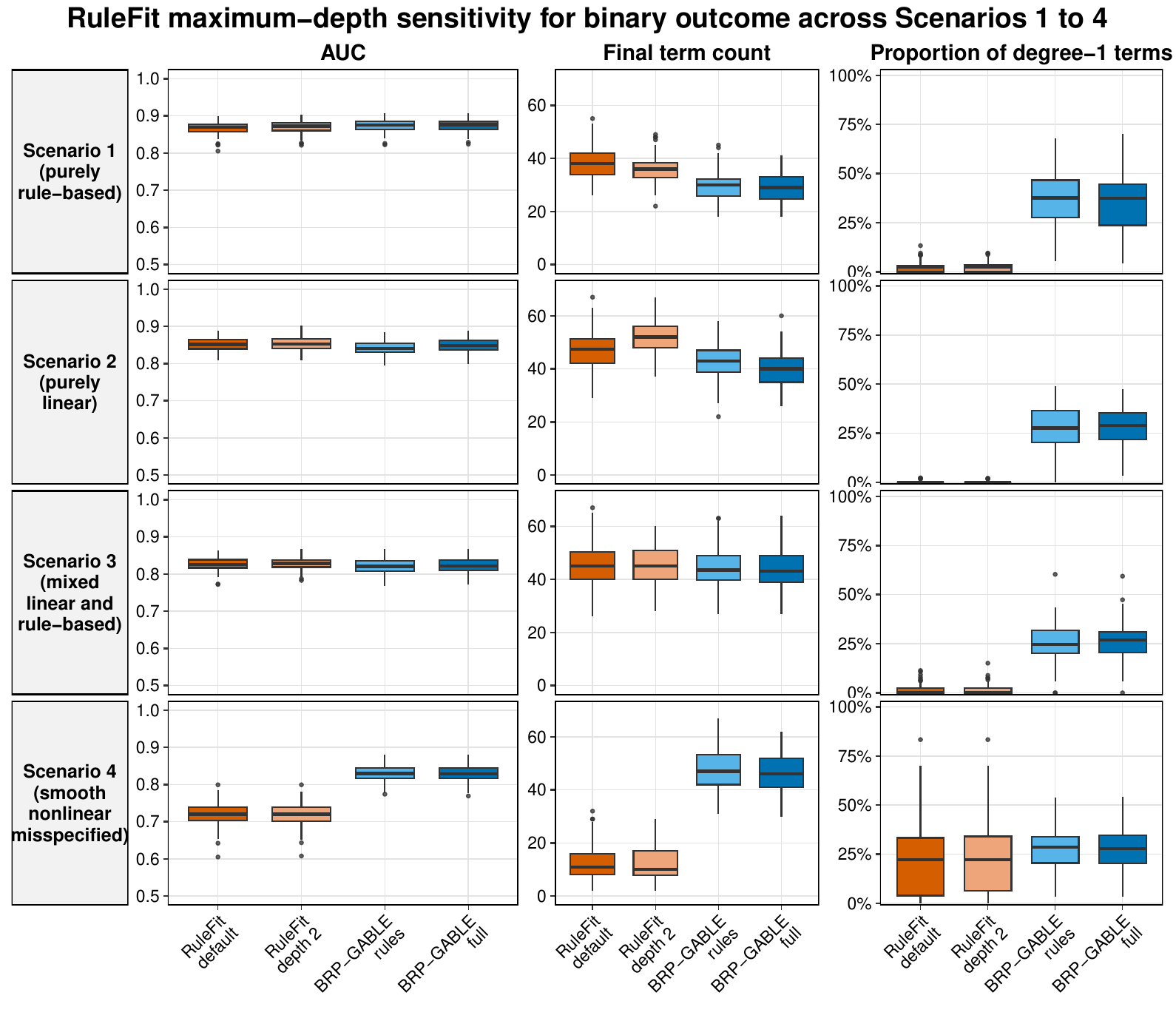}
\caption{Analysis comparing the two BRP-GABLE specifications with RuleFit under its package-default maximum depth of three and a maximum depth of two for binary outcomes.}
\label{fig:rulefit_depth_binary}
\end{figure}

\begin{figure}[htbp]
\centering
\includegraphics[width=\linewidth]{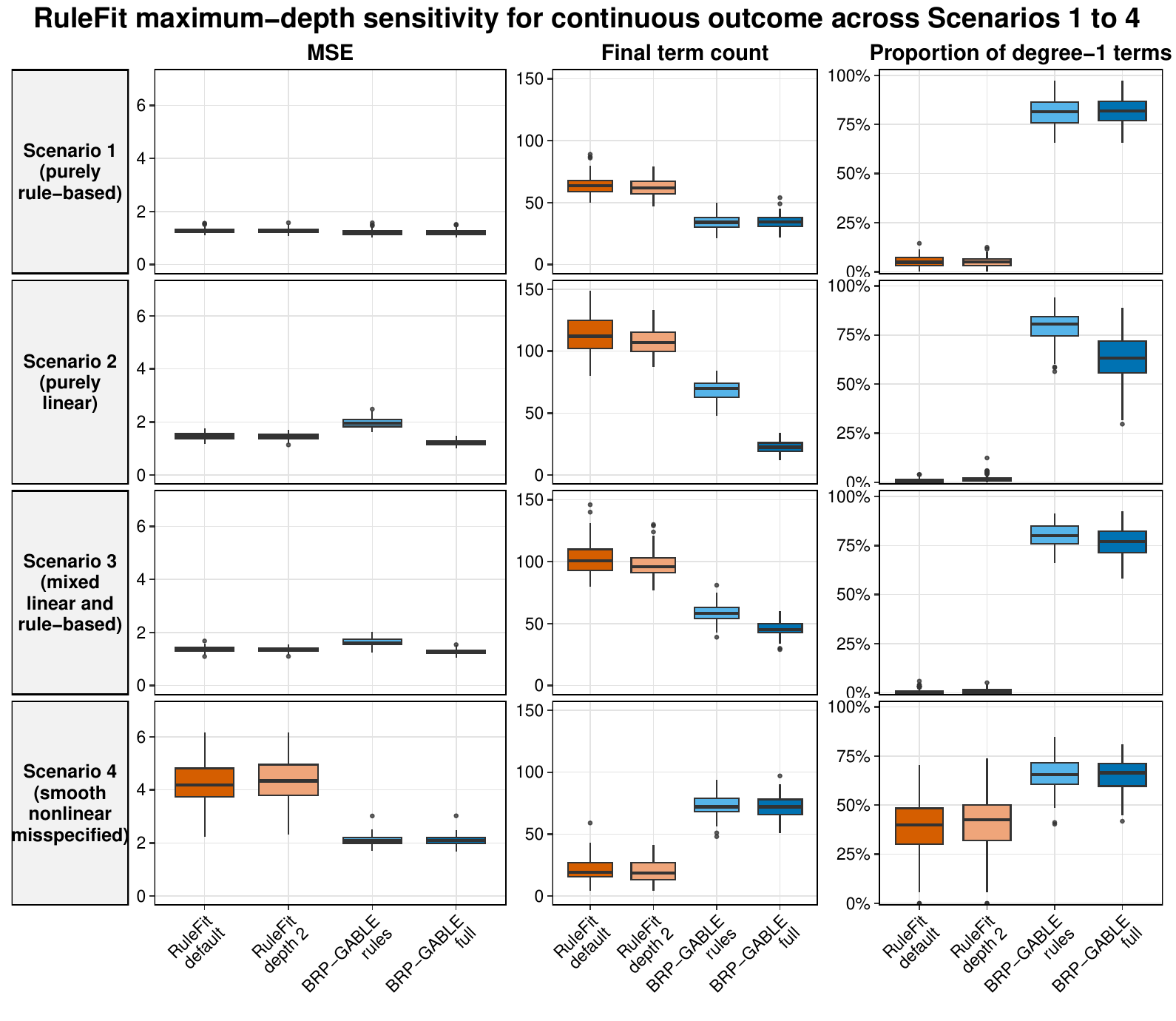}
\caption{Analysis comparing the two BRP-GABLE specifications with RuleFit under its package-default maximum depth of three and a maximum depth of two for continuous outcomes.}
\label{fig:rulefit_depth_continuous}
\end{figure}

\begin{figure}[htbp]
\centering
\includegraphics[width=\linewidth]{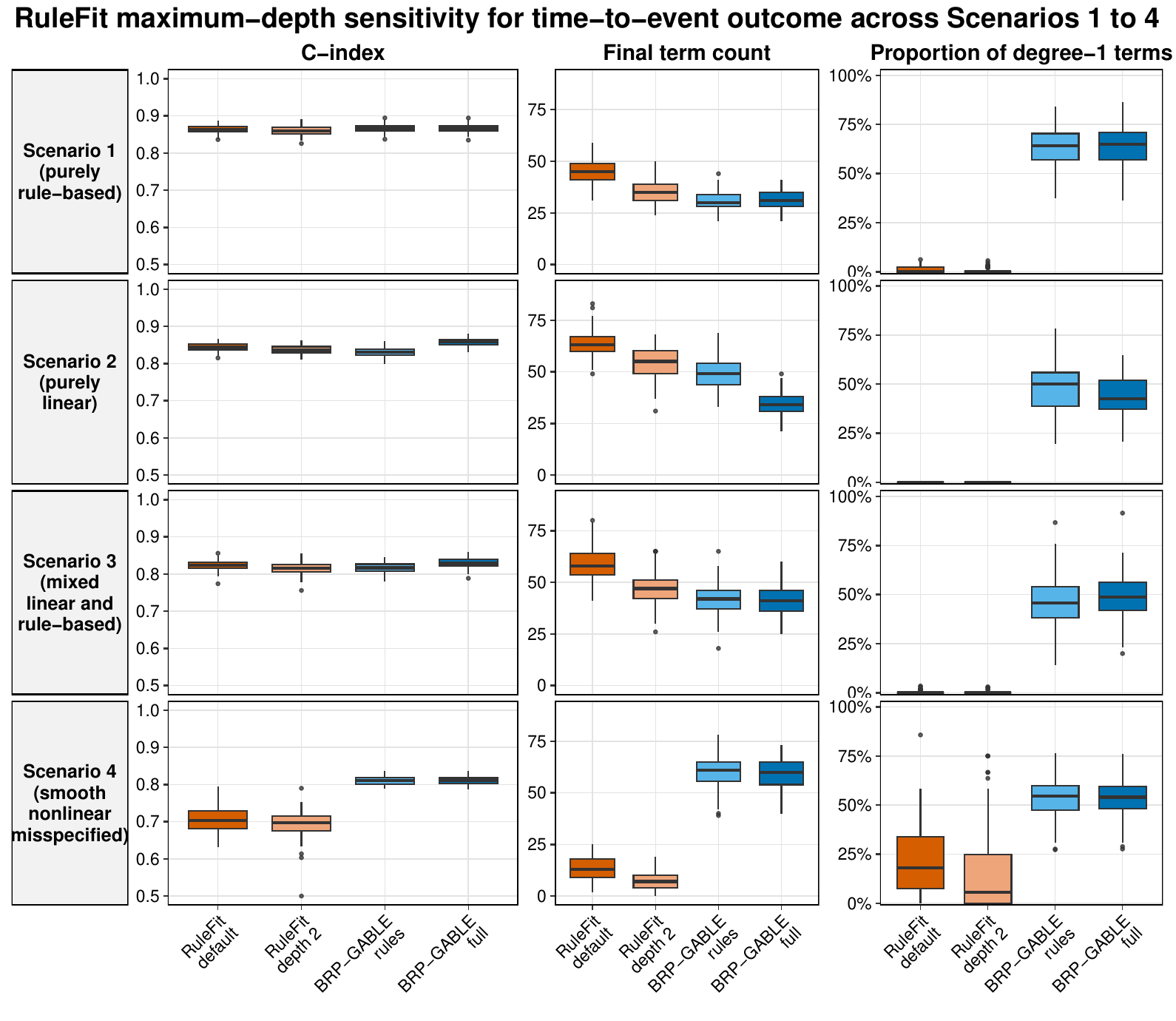}
\caption{Analysis comparing the two BRP-GABLE specifications with RuleFit under its package-default maximum depth of three and a maximum depth of two for time-to-event outcomes.}
\label{fig:rulefit_depth_survival}
\end{figure}

Figures~\ref{fig:rulefit_depth_binary}--\ref{fig:rulefit_depth_survival} present the results for the binary, continuous, and time-to-event outcomes. Restricting the maximum RuleFit depth produced little change in predictive performance for binary and continuous outcomes. For the time-to-event outcome, the depth-2 specification showed modestly lower C-index values, with the difference being more apparent in Scenarios~3 and 4. Reducing the maximum depth generally decreased the number of final RuleFit terms, although this pattern was not uniform across all settings. In Scenarios~1--3, the depth-2 RuleFit specification retained a number of final terms comparable to that of BRP-GABLE. Moreover, the proportion of degree-1 terms remained substantially lower than that of both BRP-GABLE specifications across the three outcome types. These results indicate that the main findings concerning the structural simplicity of BRP-GABLE were not explained solely by the default RuleFit setting, which permits degree-3 rules.

\clearpage

\end{document}